\documentclass[reprint, amsmath,amssymb, aps,english]{revtex4-2}

\usepackage[english]{babel}
\usepackage{bbm}
\usepackage{graphicx}
\usepackage{dcolumn}
\usepackage{bm}
\usepackage{bbm}
\usepackage{hyperref}
\usepackage{overpic}
\usepackage{xcolor}
\usepackage{rotating}
\renewcommand{\vec}[1]{\boldsymbol{#1}}

\usepackage{booktabs}
\usepackage{makecell}
\renewcommand{\[}{\begin{equation}}
\renewcommand{\]}{\end{equation}}
\newcommand{\topo}{\textrm{top}}

\begin{document}

\title{A unified field theory of topological defects and non-linear local excitations}
\author{Vidar Skogvoll}
\affiliation{PoreLab, The Njord Centre, Department of Physics, University of Oslo, P. O. Box 1048, 0316 Oslo, Norway}

\author{Jonas Rønning}
\affiliation{PoreLab, The Njord Centre, Department of Physics, University of Oslo, P. O. Box 1048, 0316 Oslo, Norway}

\author{Marco Salvalaglio}
\affiliation{Institute of Scientific Computing, TU Dresden, 01062 Dresden, Germany}
\affiliation{Dresden Center for Computational Materials Science, TU Dresden, 01062 Dresden, Germany}

\author{Luiza Angheluta}
\affiliation{PoreLab, The Njord Centre, Department of Physics, University of Oslo, P. O. Box 1048, 0316 Oslo, Norway}

\date{\today}

\begin{abstract}

Topological defects and smooth excitations determine the properties of systems showing collective order. We introduce a generic non-singular field theory that comprehensively describes defects and excitations in systems with $O(n)$ broken rotational symmetry. Within this formalism, we explore fast events, such as defect nucleation/annihilation and dynamical phase transitions where the interplay between topological defects and non-linear excitations is particularly important. To highlight its versatility, we apply this formalism in the context of Bose-Einstein condensates, active nematics, and crystal lattices. 

\end{abstract}

\maketitle

\section{Introduction}\label{Sec:Intro}

Topological defects are hallmarks of systems exhibiting collective order. 
They are widely encountered from condensed matter, including biological systems, to elementary particles, and the very early Universe~\cite{merminTopologicalTheoryDefects1979,TWBKibble_1976,michel1980symmetry,vilenkin1994cosmic,chaikin_lubensky_1995,nelson2002defects,ozawa2019topological,ardavseva2022topological}. 
The small-scale dynamics of interacting topological defects are crucial for the emergence of large-scale non-equilibrium phenomena, such as quantum turbulence in superfluids~\cite{madeira2020quantum}, spontaneous flows in active matter~\cite{alert2022active}, or dislocation plasticity in crystals~\cite{papanikolaou2017avalanches}. 
In fact, classical discrete modeling approaches such as point vortex models~\cite{eyink2006onsager} and discrete dislocation dynamics~\cite{zhouDiscreteDislocationDynamics2010} describe turbulence and plasticity in terms of the collective dynamics of topological defects as interacting charged points (in 2D) or line defects (in 3D). 
In most of these theories, the interactions of topological defects are modeled through the linear excitations that they induce in the far fields.
The physics of events on short time- and length scales, such as core energies, nucleation conditions, defect interaction, etc., are often introduced by ad-hoc rules, such as cut-off parameters, Schmidt stress nucleation criteria, and defect line recombination rules. 
However, the dynamics of these events play a vital role in the transitions between different dynamical regimes.
This is the case, for example, in stirred Bose-Einstein condensates where different superfluid flow regimes are observed depending on the size and speed of the moving obstacle~\cite{neely2010observation,kwon2015critical,aioi2011controlled,kunimi2015metastability,skaugenVortexClusteringUniversal2016,neely2013characteristics}, and where there is a subtle interplay between vortices and shock waves. 
Active nematic fluids are characterized by a dynamic transition to active turbulence at a sufficiently large activity where the spontaneous flows are sustained by the creation and annihilation of orientational defects~\cite{thampi2014vorticity,doostmohammadi2017onset}. 
During plastic deformation of polycrystals, grains are progressively fragmented, a process governed by the nucleation and patterning of dislocations \cite{zolotorevskyLargescaleFragmentationGrains2022}. 
A number of macroscopic criteria exist for the nucleation of topological defects in crystals \cite{liElasticCriterionDislocation2004,millerStressgradientBasedCriterion2004,gargStudyConditionsDislocation2015}. 
Due to the highly non-linear nature of this process, however, it still remains poorly understood.

In this paper, we present a formalism to describe the evolution of ordered systems from the dynamics of their topological defects and their interactions with smooth but localized excitations.  
The versatility of the approach allows us to gain insight into defect annihilation, the onset of collective behavior, and perspectives on defect structures. 
In particular, we apply the method to systems of increasing topological and dynamical complexity.
First, we study the motion of isolated vortices in Bose-Einstein condensates, which, in addition to confirming that the method correctly identifies topological defects and their velocities, sheds light on changes in quantum pressure arising from the interplay between phase slips and shock waves. 
For active nematics, we observe that the onset of active turbulence as a melting of periodic arches is signaled by the formation of bound dipoles of nematic defects at the core of dislocations in the nematic arches. 
Similarly, bound dipoles of phase slips are also associated with the nucleation of dislocations in a crystal lattice.

The proposed approach builds upon the classical method introduced by Halperin and Mazenko (hereafter called the HM-method) \cite{halperinStatisticalMechanicsTopological1981,mazenkoVortexVelocitiesSymmetric1997} to track and derive analytical results for topological defects.
Therefore, in Section \ref{sec:classical_topological_decription}, we begin with preliminary details of homotopy theory for topological defects and how the HM-method can be used for $O(n)$-symmetric theories to track their location and kinematics. 
In Sec.~\ref{Sec:Defect_fields}, we then develop a non-singular field theory as a generalization of the HM-method which constitutes our primary reduced defect field.
The method is then applied to the aforementioned physical systems in Secs.~\ref{Sec:BEC}-\ref{Sec:Crystals}. 
For the sake of readability, a rigorous derivation of the theoretical framework for arbitrary dimensions and details of the numerical simulations are reported in the Supplementary Notes \cite{SI}.   
Conclusions and perspectives for further study are outlined in the final section~\ref{Sec:Conclusions}.

\subsection{Classical description of topological defects}
\label{sec:classical_topological_decription}

Collective order is typically described by an order parameter field representative of symmetries and carrying information about topological defects and smooth, localized excitations.
Although the order parameters are well-established for conventional systems, one often needs to define them for more exotic systems 
~\cite{Mietke2022,monderkamp2022topological}. 
In this paper, we focus on well-known order parameters for systems with broken $O(n)$ rotational symmetries, where $n$ is the intrinsic dimension of the order parameter.

Homotopy theory provides a valuable identification and classification of topological defects~\cite{merminTopologicalTheoryDefects1979}.
The fundamental idea of homotopy theory is that the order parameter can be mapped onto a particular topological space $R$ and the \textit{homotopy} group of $R$ classify topological defects. 
For example, in the $XY$-model of ferromagnetism, and more generally for any system with $O(2)$ broken symmetry, the order parameter is mapped by a 2D unit vector $\vec u$ onto $R=\mathcal S^1$, the unit circle.
On $\mathcal S^1$, we may define classes of closed circuits (loops), where loops of the same class are \textit{homotopic}, i.e., they can be continuously deformed into each other.
These classes, together with an appropriate binary operation, define the homotopy group of $\mathcal S^1$. 
This group is isomorphic to $\mathbb Z$ under addition since the difference between two loops that are not homotopic is how many times they have looped around the circle $\mathcal S^1$.
Therefore, in regions of space where $\vec u$ is continuous and well-defined, a closed circuit $\partial \mathcal M$ in real space corresponds to a closed circuit in $\mathcal S^1$, and the topological charge $s_{\textrm{top}}$ contained in $\partial \mathcal M$ is given as an integer by the isomorphism between homotopy group of $\mathcal S^1$ and $\mathbb Z$.
This topological charge is obtained from  $\vec u=(\cos \theta,\sin \theta)$, by the contour integral 
\[
s_{\textrm{top}} = \frac{1}{2\pi} \oint_{\partial \mathcal M} d\theta,
\label{eq:integer_charge}
\]
which is invariant under any smooth deformations of $\partial \mathcal M$. 
This also implies that by shrinking $\partial \mathcal M$ down to a point and given that $s_{\textrm{top}}$ is a constant, there must be regions inside $\partial \mathcal M$ where $\vec u$ is undefined. 
These are the topological defects that have a $s_{\textrm{top}}$ charge.
Therefore, topological defects for $R=\mathcal S^1$ in 2D are points with their charge determined by corresponding loop integration. 
On the other hand, such topological defects (with $R=\mathcal S^1$) in three dimensions are \textit{lines}.

\begin{figure}[htp]
    \centering\includegraphics[width=0.48\textwidth]{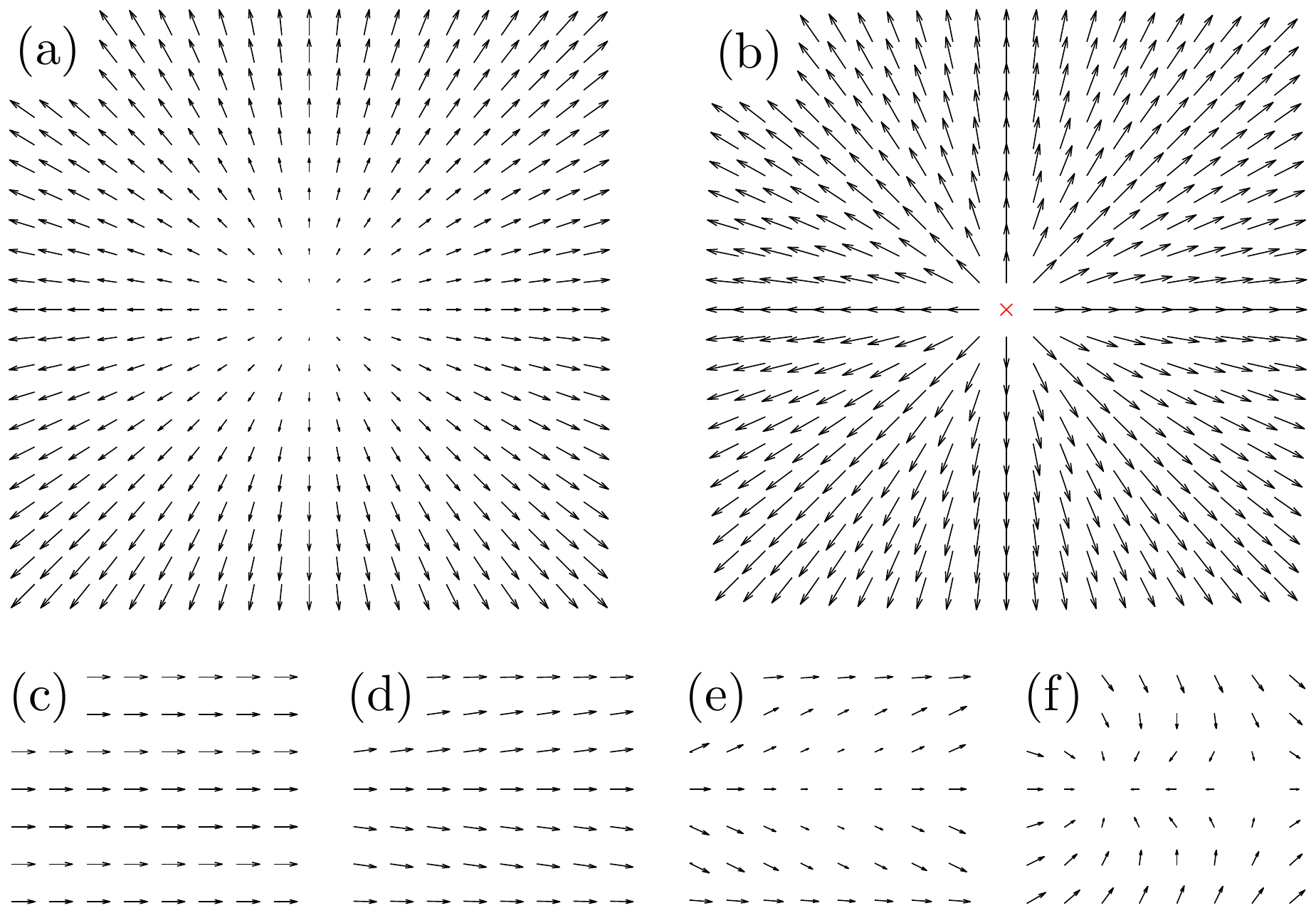}
    \caption{Different types of excitations in a 2D vector field theory. A $+1$ defect is shown in (a) the order parameter field $\vec \Psi$ and (b) in the unit vector field $\vec u=\vec \Psi/|\vec \Psi|$.
    Excitations of (c) the ground state can be categorized into (d) linear excitations with variations in the orientation of $\vec \Psi$, (e) local non-linear excitations for which also the magnitude $|\vec \Psi|$ varies and (f) topological defects.}
    \label{fig:psi_vs_n}
\end{figure}

In field theories of symmetry-breaking transitions, the ground state of the order parameter minimizes a free energy constructed from symmetry considerations. 
For broken rotational symmetries, the order parameter is a vector field $ \vec \Psi$, which in the ordered (ground) state has a constant magnitude $|\vec \Psi|=\Psi_0$, meaning that the ground state manifold is $\mathcal S^{n-1}$, where $n$ is the number of components of $\vec\Psi$. 
The link between the order parameter $\vec \Psi$, and 2D unit vector (director) field $\vec u \in \mathcal S^1$ is given by $\vec u = \vec \Psi/|\vec \Psi|$ and topological defects are located at positions where $\vec u$ is undefined, which corresponds to $|\vec \Psi|=0$ as shown in Fig.~\ref{fig:psi_vs_n} (a-b).

A description of topological defects as zeros of order parameters in $O(n)$ models and their kinematics was proposed originally by Halperin and Mazenko in the context of phase-ordering kinetics~ \cite{halperinStatisticalMechanicsTopological1981,mazenkoVortexVelocitiesSymmetric1997} and extended to systems driven out of equilibrium, such as in stirred Bose-Einstein condensation \cite{skaugenVortexClusteringUniversal2016,ronningClassicalAnalogiesForce2020,ronningNucleationKinematicsVortices2022}, 
active nematics~\cite{anghelutaRoleFluidFlow2021,ronningFlowTopologicalDefects2021}, and deformed crystals~\cite{skaugenDislocationDynamicsCrystal2018,skogvollDislocationNucleationPhasefield2021,skogvollPhaseFieldCrystal2022}.
Sticking to $O(2)$-symmetry in two dimensions and using the definition of a topological charge given in Equation \eqref{eq:integer_charge}, it is possible to express the topological defect density in terms of the zeros of the order parameter $\vec \Psi$~\cite{halperinStatisticalMechanicsTopological1981} tracked by Dirac-delta functions as
\[
\rho_{\textrm{top}}(\vec r) \equiv \sum_{\alpha} q_\alpha \delta^{(2)}(\vec r-\vec r_\alpha) = D(\vec r) \delta^{(2)}(\vec \Psi),
\label{eq:charge_density_HM}
\]
where $q_\alpha$ and $\vec r_\alpha$ are, respectively, the charge and position of the topological defect $\alpha$, $\delta^{(2)}(\vec \Psi) = \delta(\Psi_1)\delta(\Psi_2)$, and $D(\vec r)$ is the (signed) Jacobian determinant of the map $\vec \Psi$, 
\begin{equation}
    \begin{split}
D = \frac{\partial(\Psi_1,\Psi_2)}{\partial (x,y)} = &\partial_x \Psi_1 \partial_y \Psi_2 - \partial_x \Psi_2 \partial_y \Psi_1 \\
=
&\frac{1}{2}  \epsilon^{ij} \tilde \epsilon^{mn}
(\partial_i \Psi_m) (\partial_j \Psi_n),
\label{eq:D_field}
\end{split}
\end{equation}
where $\epsilon^{ij}$ are the components of the Levi-Civita tensor in real space.
The Levi-Civita tensor $\tilde \epsilon$ in order parameter space is written with a tilde to emphasize that it is contracted with the order parameter $\vec \Psi$. In the Cartesian space, both $\epsilon$ and $\tilde \epsilon$ are simply the Levi-Civita (permutation) symbols. 
Note that Equation \eqref{eq:charge_density_HM} is the usual scaling property of the delta function taking $\vec \Psi$ as input, apart from the sign of $D$ carrying information of the charge $q_\alpha$ of the topological defects. 
This result was shown in Ref. \cite{halperinStatisticalMechanicsTopological1981} by considering as explicit ansatz a negative point defect, but can, in general, be justified using differential forms. 
Nominally, the $D$ field in Equation \eqref{eq:D_field} is evaluated at the location of the defect only, because of the $\delta$-function in  Equation \eqref{eq:charge_density_HM}.

\section{Results}
\subsection{Non-singular defect fields}\label{Sec:Defect_fields}

The $\delta$-function in the topological charge density of Equation \eqref{eq:charge_density_HM} locates the topological defects at singular points where $\vec u$ is undefined. 
In $O(2)$ models, however, even though the ground state manifold is $\mathcal S^1$, the topological excitations have a finite core over which the magnitude of the order parameter goes smoothly to zero. 
This feature is also seen in physical systems, for instance, in liquid crystals, where optical retardance is an order parameter that goes to zero at the core. 
This has been used to quantify the size and structure of the defect cores in liquid crystals
~\cite{zhouFineStructureTopological2017}.
Motivated by this, we seek to generalize Equation (\ref{eq:charge_density_HM}) in a way that will avoid singularities in the resulting charge density.

\begin{figure*}
    \centering
    \includegraphics[width=\textwidth]{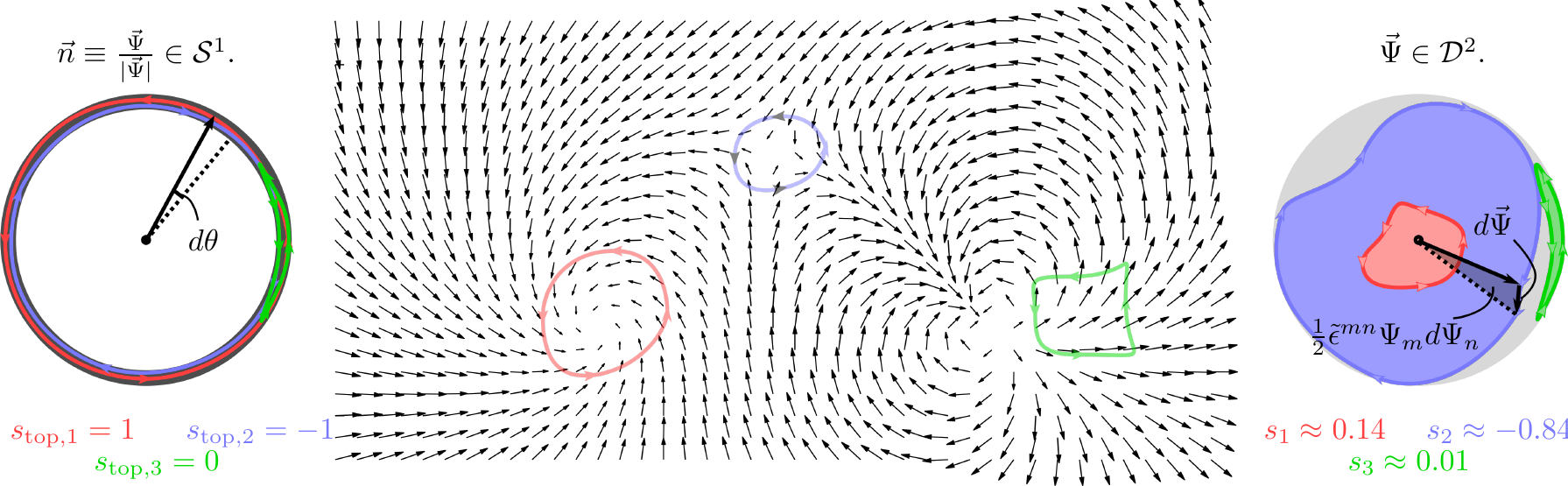}
    \caption{
    A continuous field $\vec \Psi(\vec r)$ containing defects with integer charges $+1$, $-1$ and $+2$. 
    The net integer topological charge contained in the circuits is given by the winding number of the unit vector field $\vec u$ in $\mathcal S^1$. 
    The (signed) relative area gives the value of $s$ for the circuits spanned by the order parameter $\vec \Psi$ in $\mathcal D^2$. 
    }
    \label{fig:example_of_topological_defects}
\end{figure*}

Since the equilibrium value $\Psi_0$ of $|\vec \Psi|$ is constant, the order parameter effectively resides in $\mathcal D^2$, the unit disk. 
We propose in this paper that the simplest generalization of $s_{\textrm{top}}$ is to consider the relative area of $\mathcal D^2$ swept by $\vec \Psi$ on the circuit $\partial \mathcal M$. 
During an infinitesimal displacement along $\partial \mathcal M$, $\vec \Psi$ sweeps the infinitesimal area given by half of the parallelogram spanned by $\vec \Psi$ and $d\vec \Psi$. 
This (signed) area is given by $\frac{1}{2} \tilde \epsilon^{mn} \Psi_m d\Psi_n$, see Fig.~\ref{fig:example_of_topological_defects}. 
The complete area of $\mathcal D^2$ is $\pi\Psi_0^2$, and we define the charge $s$ as the area swept by $\vec \Psi$ relative to the area of $\mathcal D^2$,
\[
s= \frac{1}{\pi \Psi_0^2} \oint_{\partial \mathcal M} \frac{1}{2} \tilde \epsilon^{mn} \Psi_m d\Psi_n,
\label{eq:continuous_charge}
\]
where $\partial \mathcal M$ is defined as in  Equation (\ref{eq:integer_charge}).
Naming $s$ a ``charge" suggests that it satisfies a global conservation law, which we shall prove shortly.
The connection between $s$ and $s_{\textrm{top}}$ is made by recognizing that  for a path $\partial \mathcal M$ in the far field of a topological defect, where $|\vec \Psi|=\Psi_0$,  $s=s_{\textrm{top}}$. 
To see this, note that if $\vec |\Psi|=\Psi_0$, then the infinitesimal area swept by $\vec \Psi$ is simply $\frac{1}{2}\Psi_0^2 d\theta$, which inserted into Equation \eqref{eq:continuous_charge} gives Equation \eqref{eq:integer_charge}.  
Closer to the core, however, the magnitude $|\vec \Psi|$ decreases and $s$ is no longer an integer, which is why the associated defect density will give information about the core extent. 
Using Green's theorem, we get 
\[
s = \frac{1}{2\pi \Psi_0^2}\oint_{\partial \mathcal M} \tilde \epsilon^{mn} \Psi_m \partial_k \Psi_n dl^k = \int_{\mathcal M} d^2 r \rho(\vec r),
\]
where $\rho(\vec r)$ is the charge density of $s$, given by
\[
\rho(\vec r) = \frac{D(\vec r)}{\pi \Psi_0^2}.
\label{eq:charge_density}
\]
Whereas $\rho_{\textrm{top}}$ describes topological defects as point singularities in the physical space, $\rho$ describes topological defects with a finite core size.

The time derivative of Equation \eqref{eq:charge_density} gives a continuity equation
\[
\partial_t \rho + \nabla \cdot \vec J = 0, 
\label{eq:rho_continuity}
\]
with the current density determined by the evolution of the order parameter 
\[J^i = -\frac{1}{\pi \Psi_0^2} \epsilon^{ij} \tilde \epsilon^{mn}(\partial_t \Psi_m) (\partial_j \Psi_n) . 
\label{eq:J_current}\] 
Thus, $\rho$ is a globally conserved quantity, and the change in $s$ contained in a circuit $\partial \mathcal M$ is given by 
\[
\partial_t s  = 
\partial_t \int_{\mathcal M} d^2 r \rho(\vec r) = \int_{\partial \mathcal M} \vec J \cdot d\vec n,\]
where $d\vec n$ is an infinitesimal surface area normal to the circuit $\partial \mathcal M$. 
Far away from defects, $|\vec \Psi| = \Psi_0 $ and the time evolution of $\vec \Psi$ is carried by its phase $\theta(\vec r,t)$ through $\vec \Psi = \Psi_0 (\cos \theta,\sin \theta)$ which can be inserted in Equation \eqref{eq:J_current} to show that $\vec J = \vec 0$.
This means that linear perturbations of the ground state, which affect the orientation of $\vec \Psi$ only, are not described by the charge density $\rho$. 
However, it describes a certain type of local non-linear perturbations, where the magnitude is affected; see Fig.~\ref{fig:psi_vs_n}(c-f). 
We will exemplify this distinction in the applications.
Due to the standard continuity form of Equation \eqref{eq:rho_continuity}, we can connect it to a velocity field $\vec v$ through the charge flux $\rho\vec v$. 
Equation \eqref{eq:rho_continuity} only determines the current $\rho\vec v$ up to an unknown divergence free contribution $\vec K$, i.e., $\vec v = \frac{1}{\rho}(\vec J + \vec K)$, where $\nabla \cdot \vec K=0$.
However, when $\rho\neq 0$, there exists a unique velocity field $\vec v^{(\vec \Psi)}$ such that the evolution of $\vec \Psi$ can be written in a generic advection form $\partial_t \vec \Psi + (\vec v^{(\Psi)} \cdot \nabla) \vec \Psi=0$, equivalently expressed as
\[
\begin{pmatrix}
\partial_t \Psi_1 \\
\partial_t \Psi_2
\end{pmatrix}
+
\begin{pmatrix}
\partial_1 \Psi_1 & \partial_2 \Psi_1 \\
\partial_1 \Psi_2 & \partial_2 \Psi_2 \\
\end{pmatrix}
\begin{pmatrix}
v_1^{(\vec \Psi)} \\ 
v_2^{(\vec \Psi)}
\end{pmatrix}=0.
\label{eq:Psi_advection_matrix_form}
\]
This equation can be inverted to uniquely determine $\vec v^{(\vec \Psi)}$ if $\textrm{det}(\partial_i \Psi_n) = D(\vec r)\neq 0$.
To find $\vec v^{(\vec \Psi)}$ where this condition holds true, i.e. the regions of interest where also $\rho(\vec r)\neq 0$ from Equation \eqref{eq:charge_density}, it is then possible to invert Equation (\ref{eq:Psi_advection_matrix_form}). However, it is easier to insert $\partial_t \vec \Psi = -(\vec v^{(\vec \Psi)} \cdot \nabla) \vec \Psi$ into the expression $\vec J/\rho$ and see that it is the solution of Equation (\ref{eq:Psi_advection_matrix_form}).  
Thus, to fix the gauge on $\vec v$, we set $\vec K=0$ to get $\vec v = \vec v^{(\Psi)}$ and find
\[
v^i = \frac{J^i}{ \rho}
=
-2 \frac{
\epsilon^{ij} 
\tilde \epsilon^{mn}
(\partial_t \Psi_m)
(\partial_j \Psi_n) 
}
{
\epsilon^{ij} 
\tilde \epsilon^{mn} 
(\partial_i \Psi_m)
(\partial_j \Psi_n)
}
,
\label{eq:charge_density_velocity}
\]
where it is implied that repeated indices are summed over independently in the numerator and denominator.
It should be noted that the velocity $\vec v$ only describes the velocity of the defect density $\rho$ and is not, in general, the same as the advection velocity of the order parameter. 
We have only shown that \textit{if} $\rho \neq 0$ in some region then \textit{it is possible} to write the evolution of $\vec \Psi$ in this way. 
If the actual evolution of $\vec \Psi$ is given as the advection $\vec v_D$ of a density field (i.e., including the term $\vec \Psi \nabla \cdot \vec v_D$), then $\vec v \neq \vec v_D$, because the compressible part of the advection will not directly translate into the motion of topological defects. 
However, if a localized topological defect moves without changing its core structure, i.e., with a frozen core, Equation (\ref{eq:charge_density_velocity}) will give this velocity in the region of the core, which we will show in Sec. \ref{Sec:BEC}.
While the expression for the current of $D$ and the velocity equation \eqref{eq:charge_density_velocity} have previously been used in the HM-method, several important distinctions can be highlighted.
Firstly, the derivation of the $\rho$ field from the redefined charge, Equation (\ref{eq:charge_density}), shows that the field carries topological information and does not only serve as auxiliary transformation determinants of $\delta$-functions. 
Secondly, the velocity field has previously only been rigorously shown to apply to topological defects. 
In contrast, this derivation also describes the velocity of $\rho$ for other non-linear excitations. 
Thirdly, the fixing of the gauge $\vec K$ has not been adequately addressed in previous works to the authors' knowledge.
While the derivation above was done for a $n=2$ order parameter in $d=2$ spatial dimensions for simplicity, topological defects exist whenever $d\geq n$.
Equation (\ref{eq:continuous_charge}) can be generalized to arbitrary dimensions by replacing the integrand with the volume of the $n$-sphere spanned by $\vec \Psi = (\Psi_1,...,\Psi_n)$ and normalizing by the volume $V_n \Psi_0^n$ of the $n$-sphere. 
We show in the Supplementary Notes~\cite{SI} the formal derivation, and here we state the result that the charge density becomes
\[
\begin{split}
&n=1: \quad\rho_i = \frac{\partial_i \Psi}{2\Psi_0}  \\
&n\geq 2: \quad \rho_{i_1 ... i_{d-n}} = \frac{D_{i_1,...i_{d-n}}}{V_n \Psi_0^n} \\
\label{eq:charge_density_generalized}
\end{split}
\]
with 
\[
D_{i_1...i_{d-n}} = \frac{1}{n!} {\epsilon^{\mu_1...\mu_n}}_{i_1 ... i_{d-n}} \tilde \epsilon^{\nu_1 ... \nu_n} (\partial_{\mu_1} \Psi_{\nu_1})...(\partial_{\mu_n} \Psi_{\nu_n}).
\]
Generalizing the derivation of the defect kinematics, we find general expressions for the reduced defect velocity field
\begin{widetext}
\[
v^{\mu_1}  = -n 
\frac{
 \delta^{[\nu_1}_{\nu_1'}\delta^{\nu_2}_{\nu_2'} ... \delta^{\nu_n]}_{\nu_n'} 
 (\partial_t\Psi_{\nu_1})(\partial^{\mu_1} \Psi^{\nu_1'})
 \prod_{l=2}^n (\partial_{\mu_l} \Psi_{\nu_l}) (\partial^{\mu_l} \Psi^{\nu_l'})
}{
\delta^{[\nu_1}_{\nu_1'}\delta^{\nu_2}_{\nu_2'} ... \delta^{\nu_n]}_{\nu_n'} 
\prod_{l=1}^n (\partial_{\mu_l} \Psi_{\nu_l})(\partial^{\mu_l} \Psi^{\nu_l'})
}, 
\label{eq:velocity_formula_generalized}
\]
\[
\textrm{Special case } n=d:  \quad
v^{\mu_1}  = 
-n\frac{
\epsilon^{\mu_1 \mu_2 ...\mu_n}\tilde \epsilon^{\nu_1...\nu_n} (\partial_t \Psi_{\nu_1}) \prod_{l=2}^n (\partial_{\mu_l} \Psi_{\nu_l})
}{
\epsilon^{\mu_1...\mu_n}\tilde \epsilon^{\nu_1...\nu_n} \prod_{l=1}^n \partial_{\mu_l} \Psi_{\nu_l}
}, 
\label{eq:velocity_formula_special_case}
\]
\end{widetext}
where $[\nu_1 \nu_2 ...\nu_n]$ is the antisymmetrization over the indices $\nu_1 \nu_2 ...\nu_n$.
Equation \eqref{eq:velocity_formula_special_case} is the special case of $n=d$, where the velocity can be written in a simpler way.
Still, Equation \eqref{eq:velocity_formula_generalized} looks complicated due to the arbitrary number of dimensions and so we have summarized the most important cases of $n\leq d\leq 3$ in Supplementary Figure 2 of the Supplementary Notes \cite{SI}.
Thus, Eqs.~\eqref{eq:charge_density_generalized} and~(\ref{eq:velocity_formula_generalized}) are the primary general expressions of the reduced defect field.
The equations generalize the description of topological defects in the HM-method to include both topological defects and non-linear excitations. 

There are two important notes to be made on the generalization beyond the case $d=n=2$. 
Firstly, for $n\geq 2$, the charge density is a rank ($d-n$) tensor that represents the defect density per $n$-dimensional volume-oriented \textit{normal} to the manifold, e.g., how the charge density on a 2D surface is expressed in terms of the \textit{normal} vector to the surface. 
The case of $n=1$ is special because densities on one-dimensional manifolds are usually expressed in terms of the density \textit{along} the manifold, i.e., the charge density per length \textit{along} the curve. 
Secondly, in the case of $n<d$, the gauge $\vec K$ cannot be uniquely determined by looking at the evolution of $\vec \Psi$ alone. Therefore, another condition is required to obtain Equation (\ref{eq:velocity_formula_generalized}). 
This condition implies that topological defects live effectively on a $d-n$ dimensional submanifold and will move perpendicular to this structure, e.g., how the motion of a line defect is given by the velocity normal to its tangent vector. 
Due to the difference in definitions of the integrals to yield the topological content, this translates to the velocity being parallel to the charge density for $n=1$ and perpendicular to it for $n\geq 2$. 
This velocity will be normal to topological structures in the case of topological lines or walls. 
While the systems of study in this manuscript exhibit ground state manifolds with $\mathcal S^1$ symmetries ($n=2$), the generalization can be directly applied to systems with $n=1$, where the defect density represents domain walls in interfacial systems such as viscous fingering \cite{nguyenPhasefieldSimulationsViscous2010}, or with $n=3$, such as the 3D Heisenberg model of ferromagnetism, where the defect density will show emergent magnetic monopoles \cite{kanazawaDirectObservationStatics2020}.
For further discussions, see the Supplementary Notes \cite{SI}.

With the method at hand, we study phenomena involving both topological charges and non-linear local excitations through the reduced defect field and the information it conveys, such as the velocity of topological defects. 
This is done by considering progressively such phenomena in three representative systems with broken $O(2)$ symmetry and featuring increasing complexity in terms of order parameters and collective behaviors. 
Both system-specific information and general behaviors will be outlined. 
As a starting point, we consider a Bose-Einstein condensate where the order parameter is isomorphic to $\vec \Psi \in \mathcal D^2$ so that the method can be directly applied.

\subsection{Defect annihilation:\\ Vortices in Bose-Einstein Condensates}\label{Sec:BEC}

\begin{figure}[ht]
    \centering
\includegraphics[width=0.49\textwidth]{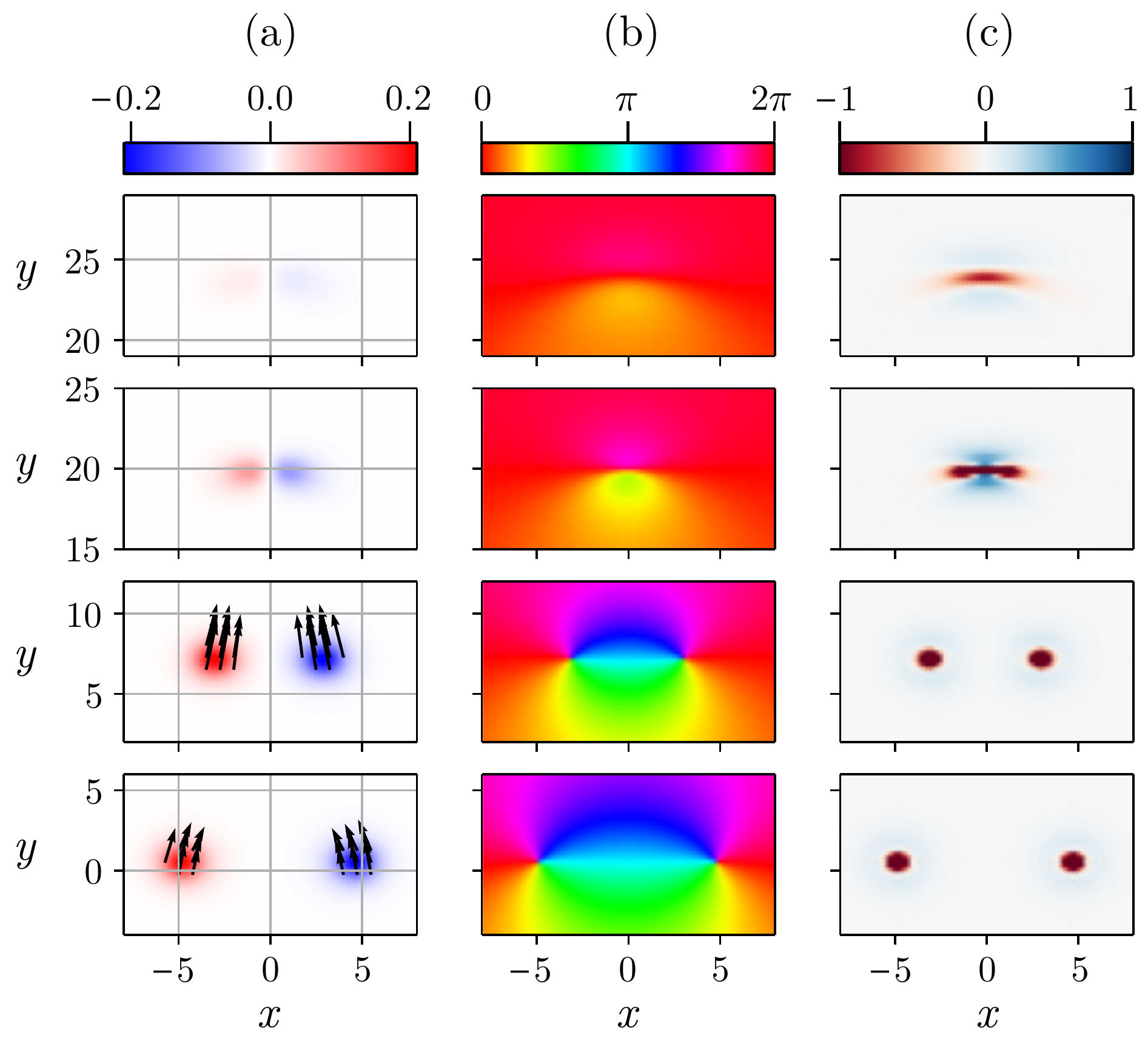}
    \caption{Annihilation of a vortex dipole in a Bose-Einstein condensate. Snapshots of (a) defect density, (b) condensate phase $arg(\psi)$ and (c) quantum pressure at different times from bottom to top:  at $t=5$, $t=60$ (before annihilation), $t=105$ (after) and $t=110$. (a) Defect velocity is included prior before annihilation. Notice in (b) the large phase gradients after the annihilation due to the induced shock-waves which can also be seen in the (c) quantum pressure profiles. The plots in column (c) have saturated colorbars because of the singular pressure at the defect core.   
    }
    \label{fig:BEC_annihilation}
\end{figure}

Within the Gross Pitaevskii theory of a superfluid Bose-Einstein condensate (BEC), the condensed bosons are described by a macroscopic wave function $\psi$, and its evolution can be described by damped Gross Pitaevskii equation~\cite{reeves2013inverse,skaugenVortexClusteringUniversal2016}
\[
\label{eq:GrossPitaeveskiiEquation}
\mathbbm i \hbar \partial_t \psi = (1-\mathbbm i \gamma) \left (
-\frac{\hbar^2}{2m} \nabla^2 + g |\psi|^2 - \mu 
\right ) \psi,
\]
where $g$ is an effective scattering parameter between condensate atoms, $\gamma>0$ is an effective thermal damping coefficient and $\mu$ is the chemical potential.
The complex condensate wavefunction $\psi$ is isomorphic to a real 2D vector order parameter $\vec \Psi = (\Psi_1,\Psi_2)$ through $\psi = \Psi_1 + \mathbbm i \Psi_2$, the norm of which is given by the absolute value $|\psi|$. 
In the equilibrium ground state, the phase of $\psi$ (and therefore the direction of $\vec \Psi$) is constant, and the magnitude is given by $|\psi|=\Psi_0= \sqrt{\mu/g}$. 
Topological defects in the orientational (unit vector) field correspond to quantized vortices captured by the charge density field 
\[
\rho^{(\psi)}(\vec r) = \frac{g D(\vec r)}{\pi\mu}.
\label{eq:BEC_density}
\]
In this context, the $D$ field (calculated from $\vec \Psi$) has the physical interpretation of the generalized superfluid vorticity
\cite{ronningNucleationKinematicsVortices2022}.
Linear perturbations of the ground state are phonons, which are characterized by traveling waves in the phase of the order parameter $\psi$, and will not be signaled in the defect density field $\rho$. 
Non-linear local perturbations, e.g., brought on by external stirring potentials or obstacles, will lead to a decrease in the magnitude of the order parameter near the obstacle \cite{astrakharchik2004motion,neely2010observation,aioi2011controlled,kunimi2015metastability}, leading to an increase in the quantum pressure, defined as 
\[
P = -\frac{\hbar^2}{2m}\frac{\nabla^2 |\psi|}{|\psi|}.
\]
Such excitations are detected by $\rho^{(\psi)}$, and mediate the nucleation or annihilation of topological defects. 
To showcase this, we simulate a BEC as dictated by Equation \eqref{eq:GrossPitaeveskiiEquation} with an initial condition featuring two vortices at $(x,y)=(\pm 5,0)$ 
Numerical details are reported in Sec. \ref{sec:appendix_numerical_methods}.
Dimensionless units are defined so that $\hbar=m=g=\mu=1$ and the damping coefficient is set to $\gamma=0.1$. 
Figure~\ref{fig:BEC_annihilation} illustrates the defect density from Equation \eqref{eq:BEC_density} during the fast event of annihilating a vortex with an anti-vortex due to a small thermal drag.  

The velocity field from Equation \eqref{eq:velocity_formula_generalized} is plotted close to vortices and shows two exciting features. 
At the beginning of the simulations ($t=5$), the non-uniform velocity over the vortex core indicates the early core deformation induced by the initial conditions. 
After this relaxation, however, vortices retain stationary or rigid cores and consistently feature a uniform velocity.
After the annihilation event, we can see traces of their diffusive cores in the excitations produced by the vortex annihilation, as seen by the quantum pressure in the system, which is shown in Fig.~\ref{fig:BEC_annihilation}(c). 
We will see in the following that similar traces appear as precursory patterns for defect nucleation.
Moreover, after having dealt with a system with only one broken symmetry, we now consider systems that have multiple rotational or translational symmetries.

\subsection{Onset of collective behavior: Active nematics}\label{Sec:Active_nematics}

In this section, we consider the case of an active nematic system. This system is peculiar as we can construct the defect density from different order parameters. By applying the proposed formalism we can investigate the transition among different regimes and the interplay among defects. Interestingly, we will show that defects in one broken symmetry are the nucleation sites of defects for a separate order. 

\begin{figure*}
    \centering
    \includegraphics[width=\textwidth]{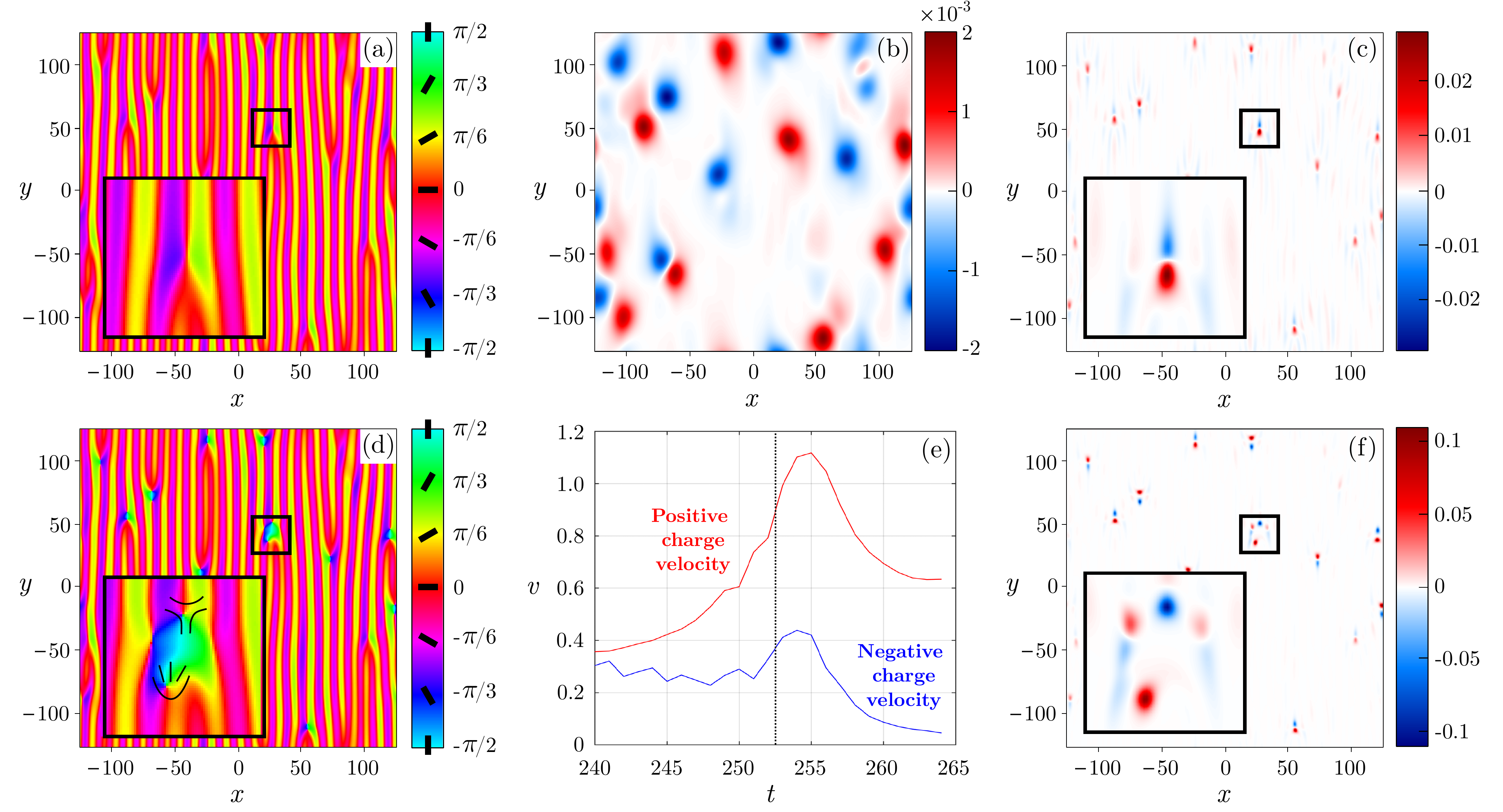}
    \caption{
    Onset of active turbulence in a nematic liquid crystal mediated by the nucleation of topological defects. (a) The angle of the nematic director at $t=240$, prior to nucleation of half-integer defects from the unstable periodic arches, and (d) at $t=260$, after nucleation. (b) The defect density $\rho^{(\eta_{\vec k})}$ at $t=240$, corresponding to the broken translational symmetry, shows the charge signature of dislocations with large core structures. The dislocation core harbors a bound dipole (inset) shown in (c) the defect density $\rho^{(Q)}$ associated to the nematic order at $t=240$ , which splits into fully formed $\pm \frac{1}{2}$ defects after nucleation as shown in (f) $\rho^{(Q)}$ at $t=260$. Panel (e) shows the speed $v = \langle |\vec v| \rangle$ of the two localized blobs in the charged defect density $\rho^{(Q)}$ around the nucleation site. 
    After the nucleation event indicated by the dashed line, these correspond to the speed of the $\pm 1/2$ defects.
    }
    \label{fig:active_nematic_order_parameters}
\end{figure*}

Within the hydrodynamic approach~\cite{marchetti2013hydrodynamics}, the nematic orientational order of active matter in two dimensions is described by a rank-2 symmetric and traceless tensor $Q$ determined by the nematic director $\mathbf n= (\cos(\theta),\sin(\theta))$
\[
Q = 
S
\begin{pmatrix}
n_1n_1 - \frac{1}{2} & n_1 n_2 \\
n_2 n_1 & n_2 n_2 - \frac{1}{2}
\end{pmatrix}
\equiv
\begin{pmatrix}
\Psi_1 & \Psi_2 \\
\Psi_2 & -\Psi_1
\end{pmatrix},
\label{eq:Q_tensor_definition}
\]
where $S$ is an order parameter which is $0$ in the disordered phase.
$Q$ is thus related to the $\mathcal D^2$ order parameter $\vec \Psi$ field by $\vec \Psi = \frac{S}{2} (\cos(2\theta),\sin(2\theta))$.
The evolution of the $Q$-tensor follows dissipative dynamics coupled with an incompressible Stokes flow with substrate friction~\cite{nejad2021memory}. 
Details on the evolution equation and its numerical method are reported in Sec \ref{sec:appendix_numerical_methods}.
The system is here initialized in a homogeneous nematic phase with small perturbations in the angle of the director field. 
These perturbations are enhanced by the active stress creating a striped phase that is further destabilized and eventually melts due to the creation of topological defects leading into active turbulence.
The ground state corresponds to a constant magnitude $|\vec \Psi| \equiv \Psi_0 = \sqrt{B}/2$ dependent on the parameter $B$, which is defined in Sec. \ref{sec:appendix_numerical_methods}.
Within the framework introduced in Sec.~\ref{Sec:Defect_fields}, this gives the following expression for the defect density
\[
\rho^{(Q)} =  \frac{4 D(\vec r)}{\pi B},
\label{eq:rho_for_nematic_order}
\]
which supports orientational defects with half-integer charge $s_{\rm top} = \pm 1/2$. 
In Fig.~\ref{fig:active_nematic_order_parameters}(a), we show the nematic orientation $\theta$ in the colorbar to emphasize the breaking of translational symmetry and the formation of a (transient) striped order.
The striped order arises from modulations in the nematic orientation which, to first order, do not change the magnitude of the order parameter $\vec \Psi$. 
Thus, these are linear perturbations not signaled by $\rho^{(Q)}$.

The inset of Fig.~\ref{fig:active_nematic_order_parameters}(a) shows a dislocation in the periodic arches in the nematic director. 
To describe these defects, we represent the parameter $\vec \Psi$ as a complex field $\psi = |\vec \Psi|e^{i\theta}$ and decompose it into a slowly-varying amplitude field of the periodic arch mode as
\[
\psi(\vec r) = \psi_0(\vec r) + \eta_{\vec k}(\vec r) e^{i\vec k\cdot \vec r} + \eta_{-\vec k}(\vec r) e^{-i\vec k\cdot \vec r}, 
\]
where $\psi_0(\vec r),\, \eta_{\vec k},\, \eta_{-\vec k}$, are slowly-varying complex fields on the length scale $a_0$ of the director field modulations.
$\vec k$ is the wave vector of the modulations which is $\vec k=\frac{2\pi}{a_0} \vec e_x$ due to the initial condition.
We can extract the complex amplitude of a $\vec k$  mode by a demodulation of $\psi$,
\[
\eta_{\vec k} = \langle \psi e^{-i\vec k\cdot \vec r} \rangle,
\label{eq:demodulation}
\]
through the convolution with a Gaussian kernel denoted by $\langle \cdot \rangle$, which filters out the small-scale variations, Equation \eqref{eq:coarse_graining_operation}. 
The modulation length scale $a_0$ and the equilibrium value $\eta_0$ of $|\eta_{\vec k}|$ are found numerically to be $a_0=10.6$ and $\eta_0=0.20$ for the given parameters.
From the order parameter $\eta_{\vec k}$, we can construct the defect density $\rho^{(\eta_{\vec k})}$ as for the complex wavefunction in the BEC. 
This field locates the dislocations from the nematic arches as shown in panel (b) at $t=240$, just prior to the nucleation of nematic defects. 

By also showing the reduced defect field $\rho^{(Q)}$ associated with the rotational symmetry (Fig.~\ref{fig:active_nematic_order_parameters}(c)), we clearly notice that each dislocation detected by $\rho^{(\eta_{\vec k})}$ is a source for the nucleation of a dipole of half-integer defects. 
The precursory pattern of the two bound defects prior to nucleation is similar to the pattern retained by the dipole annihilation  in the BEC. 
However, for active nematics, the bound state is associated with a dislocation in the periodic arches, while for BECs it is a source of quantum pressure.
We observe numerically that the melting of the smectic-like arches is mediated by the dissociation of the dislocations into dipoles of $\pm 1/2$ nematic defects. 
This occurs very fast and simultaneously at various locations, such that the system quickly transitions to active turbulence.
Notice also that the core size of the dislocations in the periodic arches is bigger than the core size of the $\pm 1/2$ nematic defects that form in the transition.
To quantify such nucleation events, we compute the defect velocity Equation \eqref{eq:charge_density_velocity} associated to the charged defect density $\rho^{(Q)}$ which is localised in well-defined blobs of opposite signs around a dislocation as illustrated in Fig.~Figure~\ref{fig:active_nematic_order_parameters}(b-c). 
By averaging the speed around these blobs, we can track the defect speed $v = \langle |\vec v| \rangle$ as function of time and show that prior to dissociation, the defects are in a bound state while afterwards they move apart as $\pm 1/2$ defects with different speeds as shown in 
Figure~\ref{fig:active_nematic_order_parameters}(e). 
Notice that the $-1/2$ defect slows down while the $+1/2$ acquires a net speed related to its self-propulsion. 

To summarise this part, our analysis offers an alternative perspective on the onset of active turbulence using the presence of competing symmetries. 
The transition to a turbulent state from a periodic arch state seems to be mediated by the dissociation of one type of topological defect into a different kind associated with changes in the global symmetries. In the following section, we study a system where the order parameters with $O(n)$-symmetry are found by decomposing a more complicated topological space.

\subsection{Defect structures: Solid Crystals}\label{Sec:Crystals}

We focus here on the study of defects and collective order in crystals. 
The ground state manifold of the crystal can be factorized in fundamental $\mathcal S^1$ spaces, which has a straightforward physical interpretation related to the crystal's Bravais reference lattice reflecting the broken translational symmetry.
As discussed below, this implies that a dislocation, i.e., a topological defect in the crystal, can be represented by bound vortices in the amplitudes of the fundamental periodic modes.
Indeed, by applying the formalism introduced in Sec.~\ref{Sec:Defect_fields}, analogies with previously discussed systems emerge, as well as peculiar features that will be discussed in detail. 

In the conserved Swift-Hohenberg modeling of crystal lattices, commonly named phase-field crystal (PFC)~\cite{Elder2002,Elder2004}, the order parameter is a weakly distorted periodic scalar field $\psi(\vec r)$, and can be approximated as
\[\label{eq:psiamp}
\psi(\vec r) = \bar \psi+ \sum_{n=1}^N \eta_n e^{i\vec q^{(n)} \cdot \vec r},
\] 
where $\bar \psi$ and $\{\eta_n \}_{n=1}^N$ are slowly varying (on the lattice unit length scale) amplitude fields, and $N$ is the number of reciprocal lattice vectors $\{\vec q^{(n)}\}_{n=1}^N$ taken into consideration. Disordered or liquid phases are described by $\eta_n (\vec r)=0$. For a perfect lattice, $\bar \psi (\vec r)= \psi_0$ and $\eta_n (\vec r) =\eta_0$ are constant, and an affine displacement $\vec r \rightarrow \vec r -\vec u$ amounts to a phase change $\eta_n=\eta_0e^{-\vec q^{(n)}\cdot\vec u}$. 
The displacement field $\vec u$ supports dislocations, which are line topological defects. 
For a path $\partial \mathcal M$ in real space circling one dislocation, the charge is given by the vector difference between the end and starting point, namely the Burgers' vector $\vec b$,  
\[
\oint_{\partial \mathcal M} d\vec u = -\vec b,
\label{eq:Burgers_vector_definition}
\]
(minus sign by convention).
The corresponding dislocation density tensor $\alpha_{ij}$ is defined through the integral of some 2D surface $\mathcal M$ bounded by $\partial \mathcal M$
\[
\int_{\mathcal M} \alpha_{ij} n^i dS = b_j,
\label{eq:dislocation_density_definition}
\]
where $\vec n$ is the normal vector to the surface element $dS$.
By multiplying Equation (\ref{eq:Burgers_vector_definition}) with a reciprocal lattice vector $\vec q^{(n)}$ of the structure, we get 
\[
\oint_{\partial \mathcal M} d(\vec q^{(n)}\cdot \vec u) = -2\pi s_n,
\]
where $s_n$ is an integer by definition of the reciprocal lattice vector. 
This shows that the phase of an amplitude $\theta_n \equiv (-\vec q^{(n)}\cdot \vec u)$ is a topological order parameter that has integer winding numbers, i.e., $\theta_n\in \mathcal S^1$. 

The amplitude $\eta_n$ acts as an order parameter in $\mathcal D^2$, i.e., $\Psi_1^{(n)}=\Re(\eta_n)$ and $\Psi_2^{(n)}=\Im(\eta_n)$. A topological description of dislocations using the HM-framework has been provided in two and three dimensions in Refs.~\cite{skaugenDislocationDynamicsCrystal2018,skogvollPhaseFieldCrystal2022}. Here, we adopt an alternative and convenient description using the charge density from Equation (\ref{eq:charge_density_generalized}), which is a vector field for 3D lattices
\[
\rho_i^{(\eta_n)} =  \frac{D_i^{(n)}}{\pi \Psi_0^2},
\]
where 
\[\vec D^{(n)}=
\begin{pmatrix}
(\partial_y \Psi_1^{(n)}) (\partial_z \Psi_2^{(n)}) - (\partial_y \Psi_2^{(n)}) (\partial_z \Psi_1^{(n)}) \\
(\partial_x \Psi_2^{(n)}) (\partial_y \Psi_1^{(n)}) - (\partial_x \Psi_1^{(n)}) (\partial_z \Psi_2^{(n)})\\
(\partial_x \Psi_1^{(n)}) (\partial_y \Psi_2^{(n)}) - (\partial_x \Psi_2^{(n)}) (\partial_y \Psi_1^{(n)})\\
\end{pmatrix}.
\]
By contracting Equation (\ref{eq:dislocation_density_definition}) with $q^j$, we can relate the dislocation density tensor with the defect charge density in a given amplitude~\cite{skogvollPhaseFieldCrystal2022}
\[
\alpha_{ij} = \frac{2d}{N\eta_0^2}\sum_{n=1}^N D_i^{(n)} q_j^{(n)},
\label{eq:dislocation_density}
\]
where $d$ is the spatial dimension. 
The amplitudes $\eta_n$ used to calculate $D^{(n)}$ are extracted from the phase-field $\psi$ as in Equation \ref{eq:demodulation}, and only the modes corresponding to the shortest reciprocal lattice vectors are used to calculate $\alpha_{ij}$.

Next, we focus on two examples to highlight insights obtained from using this approach. 
We consider the nucleation of dislocations in a square lattice from the point of view of its precursory pattern formations and quantify the dislocation core size. Then, we consider the classical inclusion problem of a rotated spherical crystal embedded in another crystal with the same lattice symmetry, to show how the surface of the inclusion changes its topology as a function of the lattice misorientation.

\begin{figure}[htp]
    \centering
    \includegraphics[width=0.48\textwidth]{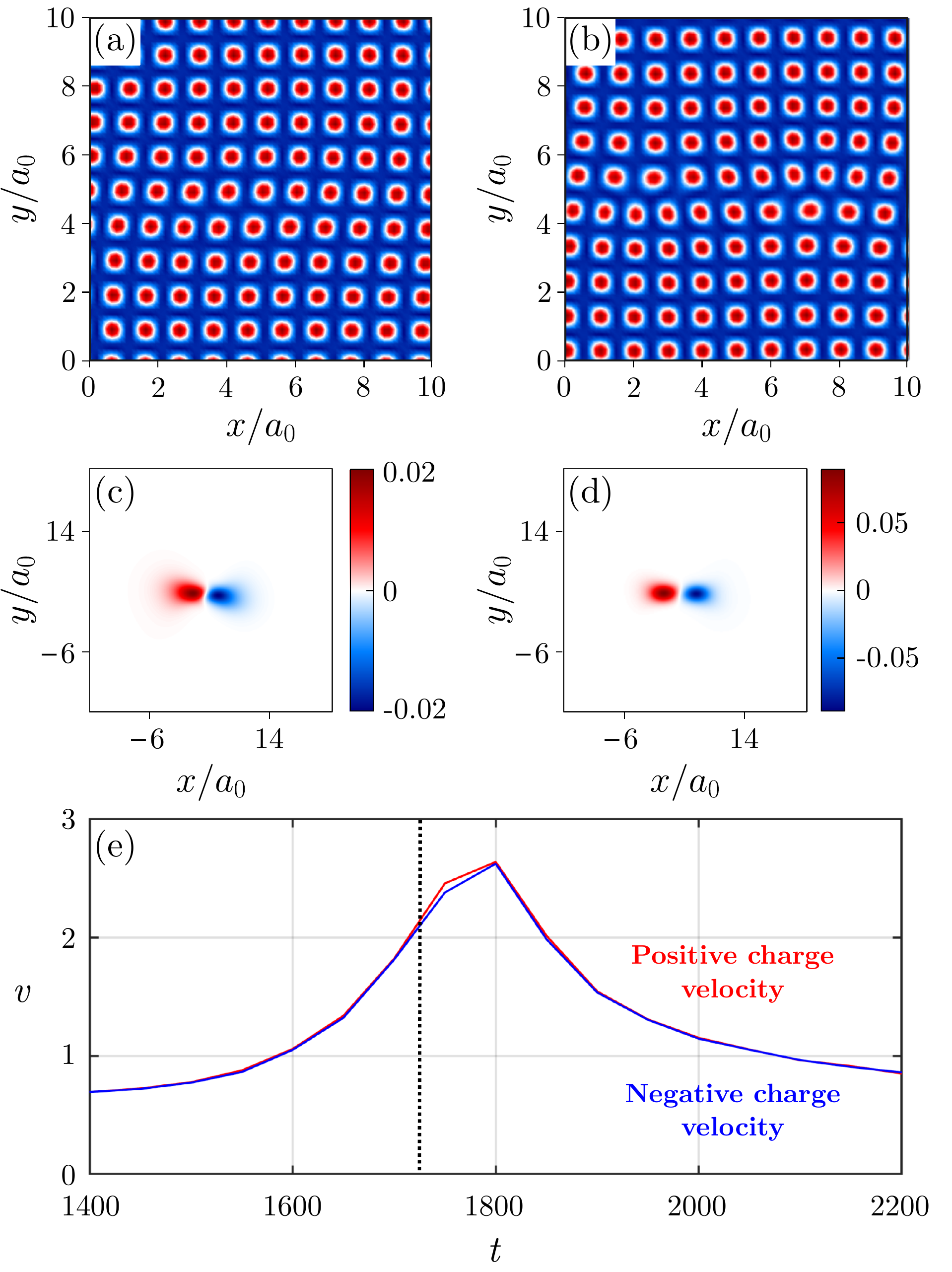}
    \caption{
    Nucleation of a dislocation dipole in a square PFC model. (a) The PFC at $t=1600$ prior to the nucleation of (b) a dislocation dipole at $t=1800$. Panels (c) and (d) show the $x$-component $\mathcal{B}_x$ of the dislocation density $\mathcal{\vec B}$ at $t=1600$ and $t=1800$, respectively. The magnitude of $\mathcal{B}_y$ is, in both cases, two orders of magnitude smaller and not shown. (e) The average speed $v=\langle |\vec v|\rangle$ at the nucleation site of positive charge ($\mathcal{B}_x>0$) and negative charge ($\mathcal{B}_x<0$) where the dashed line indicates the time of nucleation (see text). }
    \label{fig:nucleation_in_the_sq_PFC}
\end{figure}

\textit{Dislocations in 2D square lattices:}
A minimal PFC free energy which can be minimized by a square lattice reads~\cite{emdadiRevisitingPhaseDiagrams2016, skogvollStressOrderedSystems2021}
\[
F_{\psi}^{\rm sq} = \int d^2 r\left (\frac{1}{2}(\mathcal L_1 \mathcal L_2 \psi)^2 + \frac{r}{2} \psi^2 + \frac{1}{4} \psi^4 \right ),
\label{eq:Free_energy_square_PFC}
\]
where $\mathcal L_X  = X + \nabla^2$ and $r$ is a parameter. We recall that PFC energy functionals describe order-disorder (solid-liquid) phase transitions. The minimizer field $\psi$ of \eqref{eq:Free_energy_square_PFC}, for certain model parameters, has a perfect square lattice symmetry with an accurate two-mode amplitude expansion 
\[\psi = \bar \psi + \sum_{n=1}^2 \eta_n e^{i\vec q^{(n)}\cdot \vec r}
+ \sum_{n=3}^4 \eta_n e^{i\vec q^{(n)}\cdot \vec r}
+ {\rm c.c.},
\label{eq:PFC_square_reference}
\]
where $\{\vec q^{(n)} \} = \{(1,0),(0,1),(1,1),(1,-1)\}$ are the reciprocal lattice vectors of the square lattice with lengths $1$ and $\sqrt 2$. 
This sets the characteristic length $a_0=2\pi$ of the system, which is the width of the square unit cell. 
At equilibrium, the amplitude field $\eta_n$ goes to the equilibrium values $\eta_{1,2}\rightarrow A_{\textrm {sq}}$, $\eta_{3,4}\rightarrow B_{\textrm {sq}}$.
The characteristic unit of stress is given by the elastic shear modulus $\mu=16B_{\textrm{sq}}^2$~\cite{skogvollStressOrderedSystems2021}. 
 The dislocation density tensor can be factorized as $\alpha_{ij} = t_i\mathcal{B}_j$, where $\mathcal{\vec B}$ is a 2D Burgers vector density and $\vec t$ the tangent vector to the dislocation line. 
 In two dimensions, we define $\vec t$ to point out-of-plane so that the Burgers vector density is given by
 \[\mathcal{\vec B} = (\alpha_{31},\alpha_{32}),
\label{eq:coarse_2D_dislocation_density}
 \]
where $\alpha_{ij}$ can be computed by using $\vec q^{(1,2)}$.
We initiate a perfect square lattice of $101\times101$ unit cells and use the sHPFC model of Ref.~\cite{skogvollHydrodynamicPhaseField2022} to apply a local stress in the central region which causes the nucleation of a dislocation dipole.
The PFC deforms gradually, trying to account for the externally imposed stress, increasing from linear to non-linear strains until nucleation of a pure $\pm a_0\vec e_x$ dislocation dipole. 
Once formed, these dislocations move under the action of the Peach-Koehler force \cite{anderson2017}, namely they separate at large speeds due to the external stress and slow down as they reach the far-field regions of the crystal.
Simulation details are given in Sec. \ref{sec:appendix_numerical_methods}.
Figure~\ref{fig:nucleation_in_the_sq_PFC} shows the region of applied stress during the nucleation event.
Like for the nucleation of nematic defects, the nucleation is singled by a precursory localised pattern formation in the Burgers vector density, which corresponds to a bound dipole of phase slips. 
While variations only in the phase of the complex amplitudes are associated with linear elastic perturbations, non-linear elastic strains cause a decrease in the equilibrium value of the amplitudes \cite{huterNonlinearElasticEffects2016} and so produce a signal in the reduced defect density given by the expression of the dislocation density. 
Thus, the excitations visible in the dislocation density $\mathcal{\vec B}$ prior to nucleation are due to non-linear elastic strains.
From the signal profile, Fig.~\ref{fig:nucleation_in_the_sq_PFC}(c), we observe that these large non-linear elastic strains can be connected to a bound dislocation dipole.

From the defect density corresponding to $\eta_1$ for $\vec q=(1,0)$, we can also determine the average speed $v=\langle |\vec v|\rangle$ of dislocations with positive and negative charge before and after nucleation. 
The defect speed as a function of time is shown in Fig.~\ref{fig:nucleation_in_the_sq_PFC}(e). 
Like for the nucleation of defects in the active nematic, we observe a speed build up prior to nucleation succeeded by a relaxation to a constant speed. 
Unlike the $\pm 1/2$ defects in active nematics, however, both dislocations are equally mobile in this case. 

The Burgers vector density, in addition to describing the process of nucleation itself, provides us with useful information about the defect core. 
To extract the core size directly from the Burger vector density without free-tuning parameters, we consider a coarse-grained version of the PFC model, namely its amplitude expansion (APFC)~\cite{Goldenfeld2005,Athreya2006}. 
This approach gives access to phases and lattice deformation directly rather than through the demodulation of Equation \eqref{eq:demodulation}. 
It builds on the definition of a free energy functional $F_\eta$ derived from the PFC free energy $F^{\textrm{sq}}_\psi$ under the approximation of slowly-varying amplitudes. 
\begin{figure*}[htp]
    \centering
    \includegraphics[width=\textwidth]{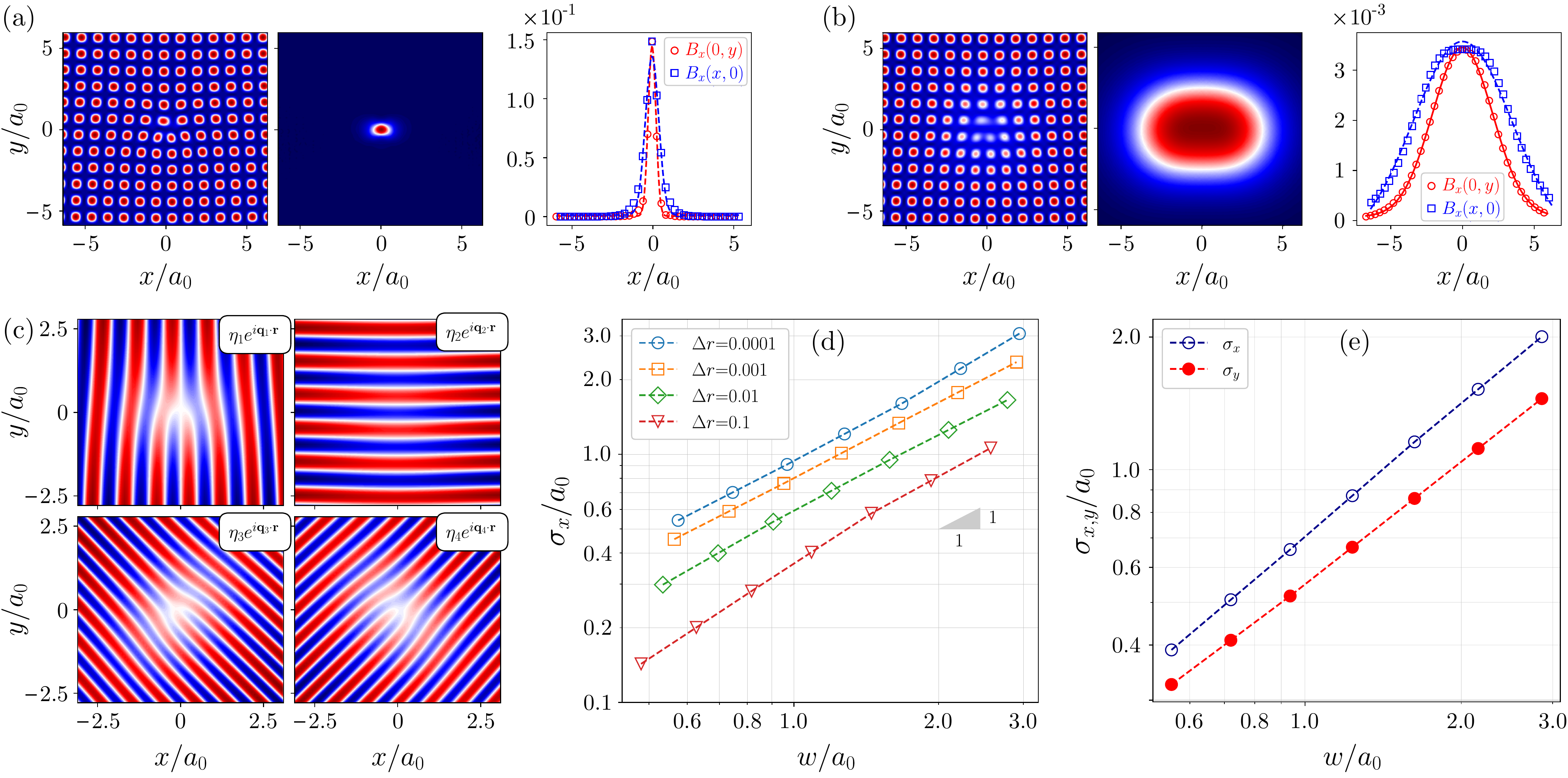}
    \caption{Dislocation core size near melting by APFC modeling. (a)--(b) Reconstructed density (left), $\mathcal{B}_x$ (center), and $\mathcal{B}_x$ along x and y direction for a relatively small and large core size, respectively obtained with (a) $\Delta r=10^{-4},\, s=3.16$ and (b) $\Delta r=10^{-1},\, s=1$, with $r_0=7.455 \cdot 10^{-2}$ the critical point. Symbols show values from APFC simulations; dashed lines correspond to Gaussian fits. The latter are exploited to quantify the size of the core in terms of the variance along x and y, namely $\sigma_x$ and $\sigma_y$. (c) Periodic modes $\eta_n e^{i\mathbf{q_n}\cdot \mathbf{r}} + \textrm{c.c.}$ for the density in panel (b). (d) Core size in terms of $\sigma_x$ as a function of the order-disorder correlation length $w$, for various values of $s$ and $r^\prime$ (the latter shown by different colors and symbols). (e) Comparison of $\sigma_x$ and $\sigma_y$ as function of $w$ for $\Delta r=0.00464$ and $s \in [10^{-1},3.16]$.}
    \label{fig:dislocore_apfc}
\end{figure*}
We simulate a square lattice hosting dislocations in a static, periodic configuration, and focus on a single defect therein.
The expression for $F_\eta$, the choice of $\mathbf{q}^{(n)}$, and details of the simulation setup are given in Sec. \ref{sec:appendix_numerical_methods}.
For the given lattice structure, the extension of its core depends on the parameters $r^\prime$ and $s$ in the free energy $F_\eta$. 
The parameter $r^\prime$ corresponds to a phenomenological temperature controlling a first-order order-disorder phase transition at $r^\prime = r_0$ with $r_0$ the critical point and ordered (disordered) phase for $r^\prime<r_0$ ($r^\prime>r_0$), and $s$ is a constant scaling the elastic moduli \cite{Elder2007,Galenko2015}. $\Delta r=r_0-r^\prime$ is referred to as the quenching depth.
These parameters affect the competition among gradient terms and the bulk energy terms in $F_\eta$. 
Fig.~\ref{fig:dislocore_apfc}(a)--(b) illustrate two different core sizes for the same dislocation obtained with different values for $r^\prime$ and $s$.
They show the reconstructed densities obtained by computing Equation \eqref{eq:psiamp} with the numerical solution for the amplitudes (first column), the Burgers vector density component $\mathcal{B}_x$ (second column), a plot of $\mathcal{B}_x(x,0)$ and $\mathcal{B}_x(0,y)$ (third column, empty symbols) with Gaussian fits (solid lines). The data fitting is obtained via $G \exp (-x^2/2\sigma_x^2-y^2/2\sigma_y^2)$ with $G$, $\sigma_x$ and $\sigma_y$ fitting parameters (dashed lines), well reproducing its shape and allowing for an estimation of the core size. 
The definition here introduced for the Burgers vector density fully characterizes the loss of coherency at the dislocation core. Importantly, it realizes a spreading of the topological charge at the core similar to non-singular continuum theories based either on regularization of singularities \cite{Cai2006} or within strain-gradient elasticity theories \cite{Lazar2005,lazar2006dislocations}. 

The amplitude expansion defined in Equation \eqref{eq:psiamp}, and thus the density field $\psi$, correspond to the sum of plane waves (Fourier modes) which are periodic stripe phases similar to the one shown in Fig.~\ref{fig:active_nematic_order_parameters}. The dislocation in the crystal then corresponds to the superposition of defects in such stripe phases. Interestingly, dislocations do not necessarily correspond to a defect for all the coupled stripe phases.
Indeed, by applying Equation \eqref{eq:integer_charge} to the phase of the amplitudes one gets $-\oint \mathbf{q}^{(n)}\cdot \mathbf{u}=2\pi\mathbf{q}^{(n)}\cdot \mathbf{b}$. 
At least for perfect dislocations, those having a translation vector of the lattice as Burgers vector, we have that $\mathbf{q}^{(n)}\cdot \mathbf{b}=0$, for some $n$. Therefore, at the dislocation core, a different ordered phase forms as some amplitudes may have non-singular phases and, in turn, do not vanish. This differs from the case of dislocations forming in pure stripe phases, e.g., in Fig.~\ref{fig:active_nematic_order_parameters}, where the single complex amplitude vanishes, pointing to a disordered phase. In Fig.~\ref{fig:dislocore_apfc}(c), the fields $\eta_n e^{i\vec q^{(n)} \cdot \vec r}$ entering the sum in Equation \eqref{eq:psiamp} are reported. Three out of four stripe phases (${n=1,3,4}$) vanish at the core, while one (${n=2}$) features a small variation of its amplitudes with no topological content. 

The defect core can then be interpreted as a transition region between two different ordered phases, one of which is present at the dislocation core only. To explore the analogy with phase interfaces, we compare its extension with the width of a solid-liquid (order-disorder) interface, $w$, which measures the correlation length for these phases. We find some analogies and differences in the dependence on the parameters entering the free energy. Traveling-wave solutions exist for solid-liquid (order-disorder) interfaces with amplitudes having hyperbolic tangent profiles, $\eta \propto (\eta_0/2)\{1-\tanh[(x-Vt)/w]\}$ with $w \propto \sqrt{g}/(1+\sqrt{1-8r^\prime/9r_0})$, $V$ the interface velocity along its normal and $g$ a parameter in the free energy which multiplies gradient terms and scales the elastic constants \cite{Galenko2015,AnkudinovTravellingWaves2020,salvalaglio2022coarse} (see also Sec. \ref{sec:appendix_numerical_methods}).
For a given set of parameters, we determine the specific amplitude profile and $w$ by fitting the result of numerical calculations with the hyperbolic tangent profile mentioned above for an interface with normal along the x-axis ($\langle10\rangle$ crystallographic direction, further details are reported in Sec. \ref{sec:appendix_numerical_methods},
Measuring the size of the dislocation core through $\sigma_x$ and $\sigma_y$ from a Gaussian fit as in Fig.~\ref{fig:dislocore_apfc}(a)-(b), we find that it scales linearly with $w$ when varying $g$, while a different scaling is observed when varying $r^\prime$, c.f. Fig.~\ref{fig:dislocore_apfc}(d).
Here $g$ is an energy scale associated with amplitudes gradients, similar to theories based on Ginzburg-Landau energy functionals \cite{salvalaglio2022coarse}. $r^\prime$, instead, affects the equilibrium values of the amplitudes, which are qualitatively different for an interface, where they all vanish in the disordered phase, and a defect, where some amplitudes are non-zero owing to a non-singular phase (see Fig.~\ref{fig:dislocore_apfc}(c)). 
Also, for $r^\prime \neq r_0$, interfaces move, which affects the width $w$ \cite{Nizovtseva2018}. 
A more detailed analysis would require finding a solution for the amplitudes' profile at defects, which goes beyond the goals of this investigation and will be addressed in future work. 

The evaluation of the Burgers vector density also allows for the characterization of anisotropies in the behavior of phases at the core as illustrated in Fig.~\ref{fig:dislocore_apfc}(d). 
$\sigma_y$/$\sigma_x\approx 0.75$ throughout the whole range of parameters investigated here as also illustrated in Fig.~\ref{fig:dislocore_apfc}(e). 
This may be ascribed to the asymmetry introduced by the specific orientation of the Burgers vector. 
We conclude that, for systems described by order parameters as in the phase-field crystal model, as well as in descriptions exploited in previous sections, the defect density may be exploited to characterize the loss of coherency at defects.

\begin{figure*}[htp]
    \centering
    \includegraphics[width=\textwidth]{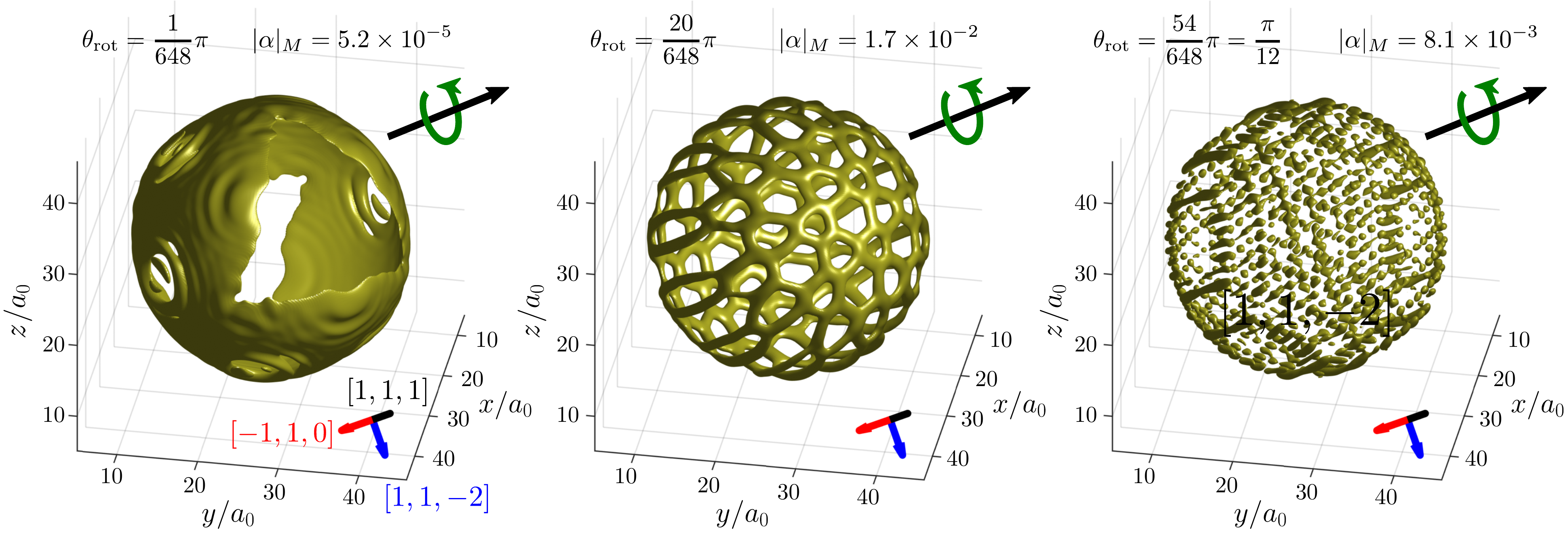}
    \caption{Rotated inclusions in the bcc PFC model. The panels show, for three representative rotation angles $\theta_{\textrm{rot}}$ the isosurface of the Frobenius norm of the coarse dislocation density tensor $|\alpha|=\sqrt{\alpha^{ij} \alpha_{ij}}$ at $50\%$ of its maximal value $|\alpha|_M = \max_{\vec r}(|\alpha|(\vec r))$, which is given in the panels. 
    }
    \label{fig:inclusion_in_bcc_PFC}
\end{figure*}

\textit{Order transition for 3D crystal inclusions: }
Like the melting of translational order in the nematic liquid crystal through the nucleation of defects in the nematic field, the global translational order in a single crystal is also destroyed under large deformations and rotations. To highlight this, we use a full 3D PFC model corresponding to a cubic lattice for which the PFC density in the one-mode approximation reads as 
\[
\psi(\vec r) = \psi_0 + \sum_{\vec q\in \mathcal R_{\rm bcc}^{(1)}} \eta_0 e^{i\vec q^{(n)} \cdot \vec r},
\]
where $\mathcal R_{\rm bcc}^{(1)}$ are the reciprocal lattice vectors of the bcc Bravais lattice with unit length~\cite{skogvollStressOrderedSystems2021}. 
This sets the length of the bcc unit cell as $a_0=2\pi\sqrt 2$. We consider spherical inclusions with radius $17a_0$, rotated at an angle $\theta_{\rm rot}$ about the $[1,1,1]$-axis.
The initial condition is relaxed by dissipative dynamics with an appropriate symmetry-conserving free energy; see further details in Sec. \ref{sec:appendix_numerical_methods}.
We choose three representative angles $\theta_{\textrm{rot}}$ and calculate the Frobenius norm $|\alpha| = \sqrt{\alpha^{ij} \alpha_{ij}}$ of $\alpha$ at each angle.
Since $|\alpha|>0$, we plot its isosurface at half its maximum value  $|\alpha|_M = \max_{\vec r}(|\alpha|(\vec r))$ in Fig.~\ref{fig:inclusion_in_bcc_PFC} for three representative misorientation angles $\theta_{\textrm{rot}}$. 
For small lattice misorientations, $|\alpha|\ll 1$, indicating only slight non-linear elastic excitations (and no fully formed dislocations) at the interface between the inclusion and the matrix.
As expected, these non-linear strains are largest in the plane perpendicular to the rotation axis, since the rotation deformation field scales with distance from the rotation axis. 
Notably, we observe a three-fold symmetry in the profile of $|\alpha|$, which can be ascribed to the underlying crystallographic orientation. 
For larger values of $\theta_{\textrm{rot}}$, the non-linear distortions increase and localize into a network of dislocations.  
Notice that such a defect network is determined directly by the Burgers vector density rather than through arbitrary reconstructions~\cite{Yamanaka2017,Salvalaglio2018}.
The description breaks down at large misorientations, as witnessed by the decrease in the magnitude of the defect density field since there is no longer a global translational order. 
Indeed, large misorientations lead to the nucleation of grain boundaries which are fully described by accounting for the bicrystallography of the two crystals meeting at the interface rather than the deformation with respect to a reference lattice \cite{HAN2018386}.
Such a regime shift echoes the onset of active turbulence in the nematic liquid, where the description in terms of the order parameter $\rho^{(\vec \eta_{\vec q})}$ also breaks down.

\section{Discussion}\label{Sec:Conclusions}

In-depth understanding and tailoring of collective behaviors require a unified description of defects associated with symmetry breaking and the non-topological excitations of ground states. 
Here, we proposed a systematic way of deriving reduced defect fields from order parameters associated with $O(n)$ broken symmetries which captures topological defects, localized non-linear excitations, and their dynamics. 
This enables the non-singular description of defects and their interaction, accounting for precursory and resulting patterns involving non-topological excitations. 
In this way, short-scale interactions between topological defects may be more accurately described, since features such as core overlap and high-energy excitations become more prominent at shorter length scales.
This paves the way for a more thorough characterization of defect interactions, particularly in cases where the defects get close or are annihilated, as in the applications shown above. 
Moreover, the proposed framework can be used to study concurrent symmetry breakings and order transitions. 
Applications to systems of general interest, such as superfluids, active nematics, and solid crystals, are shown to showcase the considered framework, while we envisage applications in many other contexts.

We have shown that the method accurately tracks topological defects since these appear as localized blobs in the defect density field. 
The associated current density and velocity field determine the kinematics of the defects, and its utility has been shown to extend beyond tracking the velocity of topological defects. 
For example, in the case of the motion of vortices in a BEC, the velocity field accounts for both the overall velocity of the defect and local variations associated with the early-stage rearrangements of the defect core evolving towards its stationary shape. Thus, the uniformity of the velocity field over the core extent tests whether the frozen-core approximation \cite{merminTopologicalTheoryDefects1979} is valid. 
For active nematics and solid crystals, the velocity formula is shown to track the dynamics of defect dipoles, during, and after the nucleation of topological defects, pointing at interesting analogies and differences between processes in different physical systems. 
The rigorous derivation of these fields given in the Supplementary Notes~\cite{SI} for any dimensions makes the equations readily applicable to tracking topological defects and localized excitations in general. 

We have found interesting features and insights about the evolution of these systems with broken symmetries. 
After the annihilation of the vortex dipole in the BEC, the remaining shock wave produces a signal in the defect density field that echoes the charge density pattern of the dipole, remnant also of other similar observations during mass-driven vortex collision \cite{richaudMassdrivenVortexCollisions2022}. 
In active nematics, the large cores of the dislocation in the translational order harbor a bound dipole of orientational defects associated with the rotational order.
This picture presents the idea of a hierarchy of topological defects, where the defects associated with one symmetry can spontaneously dissociate into stable defects for a different symmetry and melt the former ordered state. 
This is a non-equilibrium transition that echos the equilibrium Kosterlitz-Thouless transition for melting of 2D crystals via the hexatic phase~\cite{nelson1979dislocation}.

In the case of a 3D crystal, a rotated inclusion was shown to be described as a network of topological defects (dislocations) up to a point before these dissociated into other types of defects (grain boundaries) and the global orientational order was destroyed. 
The best topological description of polycrystalline materials is an open challenge, even though candidates, such as interacting disconnections \cite{HAN2018386}, exist. 
Applying this formalism to such topologies is a fascinating avenue of research.
Employing the APFC framework, where the periodic nature of crystal densities is inherently coarse-grained, we have shown that dislocation cores in Swift-Hohenberg theories emerge as transition regions from crystalline to pointwise stripe-like phases. 
When approaching the solid-liquid coexistence limit, analogies between the dislocation core size and the extensions of order-disordered interfaces have been found. 

Finally, while the whole framework is presented for systems with one broken rotational symmetry, it is a powerful tool that can be generalized to systems with multiple broken symmetries and reveal hidden hierarchies of topological defects associated with each symmetry, laying the foundation for unified theories in systems characterized by collective behaviors.

\section{Methods}
\label{sec:appendix_numerical_methods}

\subsection{\textbf{Bose-Einstein condesates}}
\label{sec:appendix_BE_condensates}

The damped Gross Pitaevskii equation, Equation \eqref{eq:GrossPitaeveskiiEquation}, is solved by using a Fourier pseudo-spectral integration scheme which is described in detail in Ref.~\cite{ronningClassicalAnalogiesForce2020}.
We use a periodic grid of size $[-32,32]\times[-32,32]$ with spatial discretization $\Delta x= \Delta y =0.25$.
To initialize the dipole we use the ansatz $\psi = \Pi_{\alpha=1}^2 \chi(|r-r_{\alpha}|)e^{i q_\alpha \theta_\alpha}$, 
where $\mathbf r_\alpha$ is the position of the vortex labeled $\alpha $, $\theta_\alpha = \arctan\left[(y-y_\alpha)/(x-x_\alpha) \right]$, and 
\begin{equation}
    \chi(r) = \begin{cases}
    r, & r< 1 \\
    1, & r> 1
    \end{cases}.
\end{equation}
This order parameter is then evolved in imaginary time, $t \rightarrow i\tau$, with $\gamma=0$ to lower the energy and find a better estimate for the core structure of the vortices we use as the initial condition. 

\subsection{\textbf{Active nematic liquid crystals}}
\label{sec:appendix_active_nematic}

The evolution of the $Q$-tensor follows dissipative dynamics coupled with an incompressible Stokes flow~ \cite{nejad2021memory}
\begin{multline}
 \partial_t Q_{ij} + \vec v \cdot \nabla Q_{ij}- Q_{ik} \Omega_{kj} + \Omega_{ik}Q_{kj}\\
 = \lambda\mathcal W_{ij} + \gamma^{-1} H_{ij}, \label{eq:Equations_of_motion_Q}
 \end{multline}
 \[
  (\Gamma - \eta \nabla^2  )v_i = \partial_j(\alpha  Q_{j i}) -\nabla p, \quad \nabla \cdot \mathbf v =0 \label{eq:Force_balance_Q},
\]
where $\vec v$ is the flow velocity that advects the nematic structure, $p$ is the fluid pressure, $\Gamma$ is the friction with a substrate, $\eta$ is the viscosity and $\alpha Q$ is the active stress.  
The vorticity tensor $2\Omega_{ij} = (\partial_i v_j - \partial_j v_i)$ rotates the nematic structure,
$\lambda$ is the flow alignment parameter which aligns the nematic orientation in the direction of shear 
\begin{equation*}
    \mathcal W_{ij} =  E_{ij} +( E_{ik}Q_{kj} +Q_{ik}E_{kj} )
    -  Q_{lk}E_{kl} (\delta_{ij} + Q_{ij}),
\end{equation*}
with the trace less strain rate $2E_{ij}=(\partial_i v_j + \partial_j v_i - \delta_{ij} \partial_k v_k)$.
The molecular field
\begin{equation}
    H_{ij} = K \nabla^2 Q_{ij} + A(B- 2Q^2_{kk}) Q_{ij}.
\end{equation}
controls the relaxation to equilibrium with $\gamma$ as the rotational diffusivity. 
We have here assumed a single Frank elastic constant $K$, treating splay and bend distortions similarly. 
The second term in the molecular field is a relaxation to a homogeneous nematic state. 
The parameter $A$ is the quench depth and $B$ sets the value of the order parameter $S_0 = \sqrt{B}$ in the homogeneous state.
We discretize the above equations on a $[-64,64]\times [-64,64]$ grid with spatial discretization $\Delta x = \Delta y =0.5$, and solve the system using pseudo-spectral methods. 
The parameters are set to $K = \Gamma = \gamma =  1$, $A=\lambda=\eta =0.5$, $B=2$ and $\alpha = -1.4$. 
The initial state is $S = \sqrt{2}$ with the angle of the director $\theta$ being uniformly distributed in the interval $(-0.05,0.05)$. 
We solve the equations for the flow field, Equation (\ref{eq:Force_balance_Q}), in Fourier space and evolve the equation for the $Q$ tensor, Equation (\ref{eq:Equations_of_motion_Q}), using the same scheme as for the BEC.

\subsection{\textbf{2D square lattice PFC}}
\label{sec:appendix_PFC_numerical_details_2D_sq_PFC}

To simulate the PFC dynamics, we use the sHPFC model proposed in Ref.~\cite{skogvollHydrodynamicPhaseField2022}, namely 
\[
\partial_t \psi =  \Gamma \nabla^2 \frac{\delta F^{\rm sq}_\psi}{\delta \psi}-\vec v \cdot \nabla \psi,
\]
coupled to a momentum equation for $\partial_t \vec v$
\[
\rho_0 \partial_t \vec v = \langle \tilde \mu_{\rm c} \nabla \psi - \nabla \tilde f  \rangle + \Gamma_S \nabla^2 \vec v + \vec f^{(\rm{ext})}.
\]
$\langle \cdot \rangle$ is a convolution with a Gaussian kernel given by 
\[
\langle {\tilde X} \rangle = \int d\vec r'\frac{\tilde X(\vec r')}{2\pi w^2} \exp\left (-\frac{(\vec r-\vec r')^2}{2w^2}\right),
\label{eq:coarse_graining_operation}
\]
 which filters out variations on length scales smaller than $w$. 
 The quench depth in Equation \eqref{eq:Free_energy_square_PFC} is set to $r=-0.3$ and the average density to $\bar \psi = -0.3$. Parameters are set to $\Gamma=1,\rho_0=\Gamma_S=2^{-6}$, and an initial velocity field $\vec v=0$. 
 We solve the system of coupled equations with a Fourier pseudo-spectral method. The spatial grid of the simulation is set to $\Delta x=\Delta y=a_0/7$. Further details can be found in Ref.~\cite{skogvollHydrodynamicPhaseField2022}.
 
In the simulation reported in Figure \ref{fig:nucleation_in_the_sq_PFC}, the perfect lattice is indented by an applied external force density given by a Gaussian profile $\vec f^{(\rm{ext})} = f_0 \frac{(y-y_0)}{a_0}  \exp\left (-\frac{(\vec r-\vec r_0)^2}{2w^2}\right )\vec e_{x}$. 
Above a critical strength $f_0=3.5\mu/a_0$ and width $w=a_0$, this force causes the nucleation of a dislocation dipole.

\subsection{\textbf{2D square lattice APFC}}
\label{sec:appendix_APFC_numerical_details_2D_sq}

The evolution of the amplitudes as delivered by the APFC model can then be directly expressed as
\begin{equation}\label{eq:dynamics_amplitudes}
\frac{\partial \eta_n}{\partial t}=-\left|\mathbf{q}^{(n)}\right|^2\frac{\delta F_\eta}{\delta \eta_n^*},
\end{equation}
with $F_\eta$ the free energy depending on $\{\eta_n\}$ that can be derived by substituting \eqref{eq:PFC_square_reference} in $F_\psi^{\rm sq}$ and integrating over the unit cell \cite{salvalaglio2022coarse}. 
By assuming constant $\bar{\psi}$ it reads
\begin{equation}\label{eq:F_eta}
F_\eta=\int d^2 r \left( g \sum_{n=1}^N |\mathcal{G}_n \eta_n|^2 + W(\{\eta_n\})  + C(\bar{\psi}) \right),
\end{equation}
with $\mathcal{G}_n=(\nabla^2+2i\mathbf{q}^{(n)}\cdot\nabla)$, $g$ a coefficient that controls elastic constants, $W(\{\eta_n\})=r^\prime\Phi/2 + (3/4) \Phi^2-(3/4) \sum_{n=1}^N |\eta_n|^4 + f^{\rm s}(\{\eta_n\})$,
$r^\prime=r+3\bar{\psi}^2$, 
$\Phi=\sum_{n=1}^N|\eta_n|^2$, and 
$f^{\rm s}(\{\eta_n\})$ a symmetry-dependent polynomial in the amplitudes. For the square symmetry as encoded in Equation \eqref{eq:Free_energy_square_PFC} and the choice $\mathbf{q}^{(1)}=(1,0)$, $\mathbf{q}^{(2)}=(0,1)$, $\mathbf{q}^{(3)}=(1,1)$, $\mathbf{q}^{(4)}=(-1,1)$ and $\{\mathbf{q}^{(n)}\}=\{-\mathbf{q}^{(n-4)}\}$ for $n=5,\dots,8$, we have $f^{\rm s}(\{\eta_n\})=2\bar{\psi}(\eta_1\eta_2\eta_3^*+\eta_1\eta_2^*\eta_4)+3(\eta_1^2\eta_3^*\eta_4+\eta_2^2\eta_3^*\eta_4^*)+\text{c.c.}$, with $\{\eta_n^*\}=\{\eta_{n-4}\}$ for $n=5,\dots,8$ as $\psi$ is a real function. Therefore, one may consider just $\eta_n$ with $n=1,\dots,4$ as variables.
$C(\bar{\psi})$ is a constant depending on $\bar{\psi}$ \cite{salvalaglio2022coarse}, set here to $\bar{\psi}=-0.3$ as set in the corresponding PFC modeling of the 2D square lattice.
$r^\prime$ corresponds to a phenomenological temperature. With $r_0$ the solid-liquid critical point, the solid crystalline phase is favored for $r^\prime < r_0$. 

We simulate a stationary system hosting dislocations with the APFC model exploiting the (FEM) numerical approach with adaptive grid refinement outlined in Refs.~\cite{salvalaglioControllingEnergy2017,praetoriusAnEfficientNumericalFramework2019}. The semi-implicit integration scheme adopted for numerical simulations can be found therein. We consider dislocations with spacing $L= 50 a_{0}$ arranged in a periodic, 2D matrix with alternating Burgers vectors $\pm a_{0}\hat{x}$. 
The system is initialized by setting the displacement field of dislocation known from classical continuum mechanics \cite{anderson2017} in the phase of amplitudes, $-\mathbf{q}^{(n)}\cdot \mathbf{u}$, and let relaxed according to the amplitudes evolution law \eqref{eq:dynamics_amplitudes}. 
We can consider a system $2L \times 2L$ by exploiting periodic boundary conditions. 

In Sec. \ref{Sec:Crystals}, we characterize the extension of the core of dislocations through the field $\mathbf{D}^{(n)}$ as entering the definition of the dislocation density tensor $\alpha$, Equation \eqref{eq:dislocation_density}. We compare the size of the defects extracted with the aid of Gaussian fits (see Figure \ref{fig:dislocore_apfc}(a)-(b)) with the extension of a solid-liquid interface, $w$, computed numerically as the average of interface width for single amplitudes. This is obtained by initializing the solid phase with a straight interface having normal along the x-axis and letting the system evolve by Equation \eqref{eq:dynamics_amplitudes} until reaching a steady state. Then, a fit of each amplitude with a function $\phi_i=\bar{A}_i[1-\tanh(x-\bar{x}_i)/\bar{w}_i]$, representing a travelling wave solution for a solid-liquid interface \cite{Galenko2015,AnkudinovTravellingWaves2020,Nizovtseva2018}, is performed with $\bar A_i$, $\bar x_i$ and $\bar w_i$ parameters and the solid-liquid interface thickness extracted as $w=\sum_{i=1}^4 \bar{w}_i/4$.

\subsection{\textbf{3D bcc lattice PFC}}
\label{sec:appendix_bcc_PFC}

Numerical simulations reported in Sec. \ref{Sec:Crystals} are obtained by solving the classical PFC  equation encoding  dissipative dynamics,
\[
\label{eq:PFC_dissipative_dynamics}
\partial_t \psi = \nabla^2 \frac{\delta F^{\rm bcc}_\psi}{\delta \psi},
\]
where $F^{\rm bcc}_\psi$ is a free energy functional that produces a stable bcc lattice, given by
\[
F^{\rm bcc}_\psi = \int d^3 r \frac{1}{2} (\mathcal L_1 \psi)^2 + \frac{r}{2} \psi^2 + \frac{1}{4} \psi^4.
\]

As parameters, we use $r=-0.3$ and $\psi_0=-0.325$ with spatial discretization $\Delta x=\Delta y= \Delta z = a_0/7$ and exploiting a Fourier pseudo-spectral integration scheme.
We consider a $51\times51\times51$ cubic crystal as matrix in which we embed a spherical inclusion with radius $17a_0$ rotated at an angle $\theta_{\rm rot}$ about the $[1,1,1]$-axis. 
This initial condition is obtained just by a rotation of grid points inside the inclusion. 
This leaves a sharp (and unphysical) interface which is regularized by letting this initial condition relax as dictated by Equation \eqref{eq:PFC_dissipative_dynamics} for $300$ time steps with $\Delta t =0.1$.

\section*{Data Availability statement}

The data will be provided by the corresponding author upon reasonable request.

\section*{Code Availability statement}

The code will be provided by the corresponding author upon reasonable request.

\section*{Acknowledgements} We are grateful to Jorge Vi\~nals for many interesting discussions.
M.S. acknowledges support from the German Research Foundation (DFG) under Grant No.~SA4032/2-1. Computing resources have been provided by the Center for Information Services and High-Performance Computing (ZIH) at TU-Dresden.

\section*{Author Contributions}
V.S: concept and theory development, simulations, data analysis, manuscript writings. 
J.R: concept and theory development, simulations, data analysis, manuscript writings.
M.S: concept and theory development, supervision, simulations, manuscript writings.
L.A: concept and theory development, supervision, manuscript writings.

\section*{Competing interests}
The authors declare that there are no competing interests.

\bibliography{Bibliography}

\section*{Figure captions}

\begin{itemize}
    \item Figure 1: Different types of excitations in a 2D vector field theory.
    \item Figure 2: A continuous field $\vec \Psi(\vec r)$ containing defects with integer charges $+1$, $-1$ and $+2$.
    \item Figure 3: Annihilation of a vortex dipole in a Bose-Einstein condensate.
    \item Figure 4: Onset of active turbulence in a nematic liquid crystal mediated by the nucleation of topological defects.
    \item Figure 5: Nucleation of a dislocation dipole in a square PFC model.
    \item Figure 6: Dislocation core size near melting by APFC modeling.
    \item Figure 7: Rotated inclusions in the bcc PFC model.
\end{itemize}

\clearpage
\onecolumngrid

\section*{Supplementary Notes}
\label{sec:appendix_generalization}

For the proofs in this section, we follow the notation conventions of Ref.~\cite{carrollSpacetimeGeometryPearson2014}.
We consider topological defects for $\mathcal S^{k}$ order parameters, which is the space consisting of $(k+1)$-dimensional unit vectors.
It is known that the $i$-th homotopy group of $\mathcal S^{k}$ is trivial for $i<k$. 
In particular, every loop ($i=1$) on the two-sphere ($k=2$) can be reduced to a point by a continuous deformation.
Thus, to get a description of the topological defects for $\mathcal S^{k}$ order parameters, we need to consider the $k$-th homotopy groups $\pi_k(\mathcal S^k) \simeq \mathbb Z$ corresponding to topological defects with integer charges.
The dimension of the defect is given by $d_{\topo} = d-(k+1)$, where $d$ is the physical space dimension. 
The homotopy classification of loops in $\mathcal S^k$ is useful beyond the direct application to models from this group because, in many systems, their order parameter space can be mapped or decomposed into products of $\mathcal S^k$ spaces. 
We define an order parameter $\vec \Psi = (\Psi_1,...,\Psi_n)$ which resides in the order parameter space $\mathcal D^n$. 
The subvolume of $\mathcal D^n$ swept by $\vec \Psi$ and $n-1$ independent variations $\{d\Psi^{(k)}\}_{k=2}^n$ is given by the (signed) volume of the $n$-dimensional cone 
\[
\frac{1}{n}\tilde  \epsilon^{\nu_1 ...\nu_n} \Psi_{\nu_1}  d\Psi_{\nu_2}^{(2)} ... d\Psi_{\nu_n}^{(n)},
\]
where $\tilde  \epsilon^{\nu_1 ...\nu_n}$  are the components of the Levi-Civita tensor in order parameter space. 
See Supplementary Figure \ref{fig:appendix_figure_generalization} for an example of $n=d=3$.
\begin{figure*}
    \centering
    \includegraphics[width=0.7\textwidth]{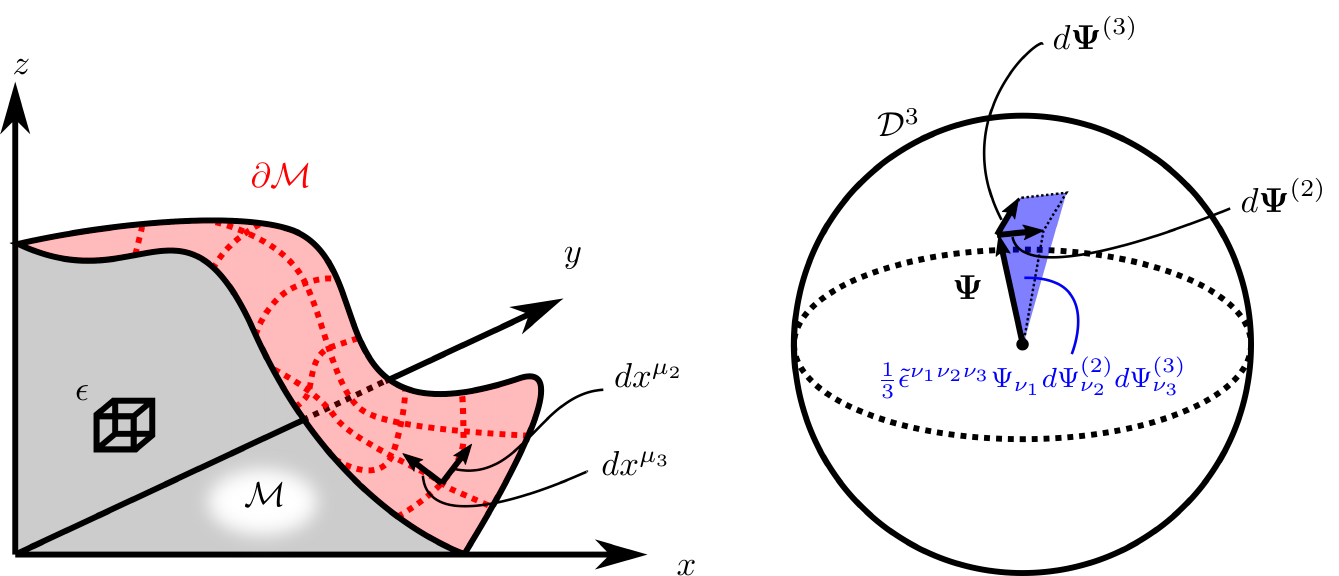}
    \caption{Real 3D space and order parameter space for a $\mathcal D^3$ order parameter. $\partial \mathcal M$ is the boundary of a 3D subvolume $\mathcal M$, on which variations along the surface ($dx^{\mu_2},dx^{\mu_3}$) lead to variations of the order parameter ($d\vec \Psi^{(2)}, d\vec \Psi^{(3)}$). 
}    \label{fig:appendix_figure_generalization}
\end{figure*}
The charge $s$ is then given as the natural generalization of Equation (4) of the main article, i.e., 
\[
s = \frac{1}{V_n \Psi_0^n} \oint_{\partial \mathcal M} \frac{1}{n} \tilde  \epsilon^{\nu_1 ...\nu_n} \Psi_{\nu_1}  d\Psi_{\nu_2}^{(2)} ... d\Psi_{\nu_n}^{(n)}.
\]
Here, $\partial \mathcal M$ is a $(n-1)$-dimensional submanifold of $\mathbb R^d$, the boundary of some $n$-dimensional submanifold $\mathcal M$, and $\{ d\vec \Psi^{(k)} \}_{k=2}^n$ are changes in $\vec \Psi$ due to displacements $dx^{\mu}$ on $\partial \mathcal M$.
Formally, we write the integrand in terms of the coordinates $\{x^{\mu}\}_{\mu=1}^d$ of $\mathbb R^d$ as follows
\[
\omega = \frac{1}{n} \tilde \epsilon^{\nu_1 ...\nu_n} \Psi_{\nu_1}  (\partial_{\mu_2} \Psi_{\nu_2}) ... (\partial_{\mu_n} \Psi_{\nu_n}) dx^{\mu_2}\otimes...\otimes dx^{\mu_n},
\]
Since $\omega_{\mu_2...\mu_n}$ is completely anti-symmetric under interchange of indices, $\omega$ is a $(n-1)$-form and we can apply Stokes' generalized theorem 
\[
\oint_{\partial \mathcal M} \omega = \int_{\mathcal M} d\omega
\]
where $d\omega$ is the exterior derivative of $\omega$, whose components are given by 
\[
(d\omega)_{\mu_1...\mu_n}=  n \partial_{[\mu_1} \omega_{\mu_2...\mu_n]}  
=
\partial_{[\mu_1}
\left (\tilde \epsilon^{\nu_1...\nu_n} \Psi_{\nu_1} (\partial_{\mu_2} \Psi_{\nu_2}) ... (\partial_{\mu_{n}]} \Psi_{\nu_n})
\right ) 
=
\tilde \epsilon^{\nu_1...\nu_n} (\partial_{\mu_1}\Psi_{\nu_1}) ... (\partial_{\mu_{n}} \Psi_{\nu_n}),
\]
where the notation $[...]$ is the antisymmetrization over free indices ($\mu_1...\mu_n$).
Thus, we get
\[
s = \frac{1}{V_n\Psi_0^n} \int_{\mathcal M} 
\tilde \epsilon^{\nu_1...\nu_n} (\partial_{\mu_1}\Psi_{\nu_1}) ... (\partial_{\mu_{n}} \Psi_{\nu_n})
dx^{\mu_1}\otimes ... \otimes dx^{\mu_n}.
\label{eq:general_charge_after_stokes}
\]
In the case of $n=1$, for which $V_1=2$, integrals over $\mathcal M$ are typically evaluated in this way, i.e., as line integrals over the one-dimensional manifold $\mathcal M$. 
We then immediately recover the defect density for $n=1$, given by 
\[
\boxed{
\rho_\mu = \frac{\partial_\mu \Psi}{2\Psi_0}
\qquad (n=1)
.
}
\label{eq:DefectDensityn1}
\]
For higher values of $n$, however, one evaluates the integral in coordinates on the manifold $\mathcal M$. 
Thus, at each point, we choose $d-n$ orthogonal unit vectors $\{\vec n_{(k)}\}_{k=1}^{d-n}$ normal to the manifold $\mathcal M$ and introduce local coordinates $\{y^i\}_{i=1}^{n}$ on $\mathcal M$.  
Expressed in these coordinates, we have 
\begin{multline}
s = \frac{1}{V_n\Psi_0^n} \int_{\mathcal M} 
\tilde \epsilon^{\nu_1...\nu_n} (\partial_{\tau_1}\Psi_{\nu_1}) ... (\partial_{\tau_{n}} \Psi_{\nu_n})
dy^{\tau_1}\otimes ... \otimes dy^{\tau_n}\\
= \frac{1}{V_n\Psi_0^n} \int_{\mathcal M} 
\tilde \epsilon^{\nu_1...\nu_n} (\partial_{\tau_1}\Psi_{\nu_1}) ... (\partial_{\tau_{n}} \Psi_{\nu_n})
\delta^{\tau_1}_{\kappa_1} ... \delta^{\tau_n}_{\kappa_n}
dy^{\kappa_1}\otimes ... \otimes dy^{\kappa_n},
\end{multline}
where $\tau$-indices iterate from $1$ to $n$. 
Now, we invoke the identity
\[
n! \delta^{[\tau_1}_{\kappa_1} ... \delta^{\tau_n]}_{\kappa_n}
=
\hat \epsilon^{\tau_1...\tau_n} 
\hat \epsilon_{\kappa_1...\kappa_n } 
\]
where $\hat \epsilon$ is the Levi-Civita tensor on $\mathcal M$, i.e., the induced volume element from $\mathbb R^d$. 
Using the fact that the integrand is already anti-symmetric in $\tau$-indices, we can replace $\delta^{\tau_1}_{\kappa_1} ... \delta^{\tau_n}_{\kappa_n} \rightarrow \delta^{[\tau_1}_{\kappa_1} ... \delta^{\tau_n]}_{\kappa_n}$ and get 
\begin{multline}
s = \frac{1}{V_n\Psi_0^n} \int_{\mathcal M} 
\tilde \epsilon^{\nu_1...\nu_n}
 (\partial_{\tau_1}\Psi_{\nu_1}) ... (\partial_{\tau_{n}} \Psi_{\nu_n})
 \frac{1}{n!} 
\hat \epsilon^{\tau_1...\tau_n}
\hat \epsilon_{\kappa_1...\kappa_n } 
dy^{\kappa_1}\otimes ... \otimes dy^{\kappa_n} \\
=
\frac{1}{V_n\Psi_0^n} \int_{\mathcal M} 
\frac{1}{n!} 
\hat \epsilon^{\tau_1...\tau_n} 
\tilde \epsilon^{\nu_1...\nu_n} (\partial_{\tau_1}\Psi_{\nu_1}) ... (\partial_{\tau_{n}} \Psi_{\nu_n})
\hat \epsilon
\end{multline}
Finally, we want to express the integrand in terms of coordinates of $\mathbb R^d$. 
Using that $\hat \epsilon$ expressed in these coordinates is 
\[
\hat \epsilon^{\mu_1...\mu_n} = N^{i_1...i_{d-n}} {\epsilon_{i_1...i_{d-n}}}^{{\mu_1...\mu_n}} 
\]
where $N^{i_1...i_{d-n}} \equiv n^{i_1}_{(1)}...n^{i_{d-n}}_{(d-n)}$, we get
\[
s=
\frac{1}{V_n\Psi_0^n} 
\int_{\mathcal M} 
\frac{1}{n!} 
{\epsilon^{\mu_1...\mu_n}}_{i_1...i_{d-n}}
\tilde \epsilon^{\nu_1...\nu_n} (\partial_{\mu_1}\Psi_{\mu_1}) ... (\partial_{\tau_{n}} \Psi_{\nu_n})
N^{i_1...i_{d-n}} 
\hat \epsilon
\]
Identifying $N^{i_1...i_{d-n}} 
\hat \epsilon$ as the oriented volume element of $\mathcal M$, we identify the defect density as
\[
\boxed{
\rho_{i_1...i_{d-n}} = \frac{D_{i_1...i_{d-n}}}{V_n \Psi_0^n}
\qquad (n\geq 2)
}
\label{eq:appendix_charge_density_generalized}
\]
where 
\[
D_{i_1...i_{d-n}} = \frac{1}{n!}
{\epsilon^{ \mu_1...\mu_n}}_{i_1...i_{d-n}}
\tilde \epsilon^{\nu_1...\nu_n} (\partial_{\mu_1}\Psi_{\nu_1}) ... (\partial_{\mu_{n}} \Psi_{\nu_n}).
\]

We now turn to finding the general equation for the velocity of the defect density. 
To differentiate Equation \eqref{eq:appendix_charge_density_generalized} with respect to time, consider
\begin{multline}
\partial_t D_{i_1...i_{d-n}} = \partial_t \left (\frac{1}{n!}
{\epsilon^{ \mu_1...\mu_n}}_{i_1...i_{d-n}}
\tilde \epsilon^{\nu_1...\nu_n} (\partial_{\mu_1}\Psi_{\nu_1}) ... (\partial_{\mu_{n}} \Psi_{\nu_n}) \right )\\
=
\frac{1}{n!}
{\epsilon^{ \mu_1...\mu_n}}_{i_1...i_{d-n}}
\tilde \epsilon^{\nu_1...\nu_n}
\left (
(\partial_{\mu_1}\partial_t\Psi_{\nu_1}) ... (\partial_{\mu_{n}} \Psi_{\nu_n}) + ... + (\partial_{\color{red}\mu_1}\Psi_{\color{blue}\nu_1}) ... (\partial_{\color{red}\mu_{n}} \partial_t \Psi_{\color{blue}\nu_n})
\right ) \\
=
\frac{1}{n!}
{\epsilon^{ \mu_1...\mu_n}}_{i_1...i_{d-n}}
\tilde \epsilon^{\nu_1...\nu_n}
\left (
(\partial_{\mu_1}\partial_t\Psi_{\nu_1}) ... (\partial_{\mu_{n}} \Psi_{\nu_n}) + ... (\partial_{\mu_{n}} \Psi_{\nu_n}) + ... + (\partial_{\color{red}\mu_n}\Psi_{\color{blue}\nu_n}) ... (\partial_{\color{red}\mu_{1}} \partial_t \Psi_{\color{blue}\nu_1})
\right )  \\
=
\frac{1}{(n-1)!}
{\epsilon^{ \mu_1...\mu_n}}_{i_1...i_{d-n}}
\tilde \epsilon^{\nu_1...\nu_n}
(\partial_{\mu_1}\partial_t\Psi_{\nu_1}) ... (\partial_{\mu_{n}} \Psi_{\nu_n}) \\
=
\partial_{\mu_1} 
\left (
\frac{1}{(n-1)!}
{\epsilon^{ \mu_1...\mu_n}}_{i_1...i_{d-n}}
\tilde \epsilon^{\nu_1...\nu_n}
(\partial_t\Psi_{\nu_1}) ... (\partial_{\mu_{n}} \Psi_{\nu_n})
\right ),
\end{multline}
where, in going from line 2 to 3, we have used that due to the contraction with both the anti-symmetric Levi-Civitas, the terms are invariant under simultaneously interchanging $\mu_{k}\leftrightarrow\mu_{k'}$ and $\nu_{k}\leftrightarrow\nu_{k'}$, so that we can write every term in the parenthesis like the first.
In going from line 4 to 5, we have used that the contraction with ${\epsilon^{ \mu_1...\mu_n}}_{i_1...i_{d-n}}$ ensures that only the first term survives when applying the product rule.
Thus, we find $\partial_t \rho_{i_1 ... i_{d-n}} + \partial_{\mu_1} {J^{\mu_1}}_{i_1 ... i_{d-n}}=0$, where
\[
{J^{\mu_1}}_{i_1...i_{d-n}} = \frac{-1}{V_n \Psi_0^n(n-1)!}
{\epsilon^{ \mu_1 \mu_2...\mu_n}}_{i_1...i_{d-n}}
\tilde \epsilon^{\nu_1...\nu_n}
(\partial_t\Psi_{\nu_1})(\partial_{\mu_2}\Psi_{\nu_2}) ... (\partial_{\mu_{n}} \Psi_{\nu_n}).
\label{eq:generalized_current}
\]
Like for the $d=n=2$ case, we want to identify this expression with the density current $v^{\mu_1} \rho_{i_1...i_{d-n}} $.
They are related up to a divergence free contribution $\partial_{\mu_1} {K^{\mu_1}}_{i_1...i_{d-n}}=0$, so
\[
v^m\rho_{i_1...i_{d-n}} 
=
{J^{\mu_1}}_{i_1...i_{d-n}} + {K^{\mu_1}}_{i_1...i_{d-n}}.
\]
In the $d=n=2$ case, the charge density on the left-hand side had no free indices and so we could simply divide by the charge density to solve for $\vec v$. 
In the general case, however, we project the equation by contracting with $\frac{\rho^{i_1...i_{d-n}}}{|\rho|^2}$, where $|\rho| = \sqrt{\rho^{i_1...i_{d-n}}\rho_{i_1...i_{d-n}}}$ is the Frobenius norm, and get
\[
v^\mu_1 = \frac{\rho^{i_1...i_{d-n}}{J^m}_{i_1...i_{d-n}}}{|\rho|^2} + \frac{\rho_{i_1...i_{d-n}}{K^{\mu_1}}_{i_1...i_{d-n}}}{|\rho|^2}.
\label{eq:velocity_after_contraction}
\]
In order to fix the gauge ${K^{\mu_1}}_{i_1...i_{d-n}}$, we look at the evolution of the order parameter $\vec \Psi$ as advected by a velocity field $\vec v_{(\Psi)}$
\[
\partial_t \Psi_n + v_{(\vec \Psi)}^{\mu_1} \partial_{\mu_1} \Psi_n = 0.
\label{eq:appendix_advection_of_psi}
\]
These are $n$ linearly independent linear equations to determine $d$ components of the velocity $\vec v^{(\Psi)}$. 
If $n<d$, it is under-determined and therefore $d-n$ additional equations are needed to determine $v^{\mu_1}$ uniquely.
We define
\begin{multline}
v^{\mu_1}_{\textrm{candidate}} \equiv 
\frac{\rho^{i_1...i_{d-n}}{J^{\mu_1}}_{i_1...i_{d-n}}}{|\rho|^2} \\
=
\frac{
\frac{1}{n!}
{\epsilon_{ \mu_1'...\mu_n'}}^{i_1...i_{d-n}}
\tilde \epsilon_{\nu_1'...\nu_n'} (\partial_{\mu_1'}\Psi^{\nu_1'}) ... (\partial_{\mu_{n}'} \Psi^{\nu_n'}) 
\frac{-1}{(n-1)!}
{\epsilon^{ \mu_1 \mu_2...\mu_n}}_{i_1...i_{d-n}}
\tilde \epsilon^{\nu_1...\nu_n}
(\partial_t\Psi_{\nu_1})(\partial_{\mu_2}\Psi_{\nu_2}) ... (\partial_{\mu_{n}} \Psi_{\nu_n})
}{
|D|^2
}\\
=
-\frac{ 
(d-n)!
\delta^{\mu_1}_{[\mu_1'} \delta^{\mu_2}_{\mu_2'} ... \delta^{\mu_n}_{\mu_n']}
\tilde \epsilon_{\nu_1'...\nu_n'} (\partial_{\mu_1'}\Psi^{\nu_1'}) ... (\partial^{\mu_{n}'} \Psi_{\nu_n'}) 
\tilde \epsilon^{\nu_1...\nu_n}
(\partial_t\Psi_{\nu_1})(\partial_{\mu_2}\Psi_{\nu_2}) ... (\partial_{\mu_{n}} \Psi_{\nu_n})
}{
 (n-1)! |D|^2
}
\\
=
-\frac{ (d-n)!
 n! \delta^{[\nu_1}_{\nu_1'} ... \delta^{\nu_n]}_{\nu_n'}  (\partial^{\mu_1}\Psi^{\nu_1'}) ... (\partial^{\mu_{n}} \Psi^{\nu_n'}) 
(\partial_t\Psi_{\nu_1})(\partial_{\mu_2}\Psi_{\nu_2}) ... (\partial_{\mu_{n}} \Psi_{\nu_n})
}{
 (n-1)! |D|^2
}
\\
=
-\frac{(d-n)!n}{|D|^2}
 \delta^{[\nu_1}_{\nu_1'} ... \delta^{\nu_n]}_{\nu_n'} 
 (\partial_t\Psi_{\nu_1})(\partial^{\mu_1} \Psi^{\nu_1'})
 \prod_{l=2}^n (\partial_{\mu_l} \Psi_{\nu_l}) (\partial^{\mu_l} \Psi^{\nu_l'})
\label{eq:appendix_a_solution_of_advection}
\end{multline}
Calculating $|D|^2$ gives 
\begin{multline}
|D|^2
=
\frac{1}{n!}
{\epsilon^{ \mu_1...\mu_n}}_{i_1...i_{d-n}}
\tilde \epsilon^{\nu_1...\nu_n} (\partial_{\mu_1}\Psi_{\nu_1}) ... (\partial_{\mu_{n}} \Psi_{\nu_n})
\frac{1}{n!}
{\epsilon_{ \mu_1'...\mu_n'}}^{i_1...i_{d-n}}
\tilde \epsilon_{\nu_1'...\nu_n'} (\partial^{\mu_1}\Psi_{\nu_1'}) ... (\partial^{\mu_n} \Psi_{\nu_n'}) \\
=
\frac{1}{(n!)^2} 
n! (d-n)! \delta^{[\mu_1}_{\mu_1'}...\delta^{\mu_n]}_{\mu_n'}
\tilde \epsilon^{\nu_1...\nu_n} (\partial_{\mu_1}\Psi_{\nu_1}) ... (\partial_{\mu_{n}} \Psi_{\nu_n})
\tilde \epsilon_{\nu_1'...\nu_n'} (\partial^{\mu_1'}\Psi_{\nu_1'}) ... (\partial^{\mu_n'} \Psi_{\nu_n'}) \\
=(d-n)! \delta^{[\nu_1}_{\nu_1'} ... \delta^{\nu_n]}_{\nu_n'} 
\prod_{l=1}^n (\partial_{\mu_l} \Psi_{\nu_l})(\partial^{\mu_l} \Psi^{\nu_l'}).
\end{multline}
which gives
\[
v^{\mu_1}_{\textrm{candidate}}
=
-n 
\frac{
 \delta^{[\nu_1}_{\nu_1'} ... \delta^{\nu_n]}_{\nu_n'} 
 (\partial_t\Psi_{\nu_1})(\partial^{\mu_1} \Psi^{\nu_1'})
 \prod_{l=2}^n (\partial_{\mu_l} \Psi_{\nu_l}) (\partial^{\mu_l} \Psi^{\nu_l'})
}{
\delta^{[\nu_1}_{\nu_1'} ... \delta^{\nu_n]}_{\nu_n'} 
\prod_{l=1}^n (\partial_{\mu_l} \Psi_{\nu_l})(\partial^{\mu_l} \Psi^{\nu_l'})
}
\]
where it is understood that the repeated indices are summed over independently in the numerator and denominator. 
\newcommand{\name}[2]{\overset{\underbrace{{#2}}}{#1}}
By inserting this expression in the LHS of  Equation (\ref{eq:appendix_advection_of_psi}) after multiplying by the denominator, we get
\[
\name{(\textrm{Mercury})}{\delta^{[\nu_1}_{\nu_1'} ... \delta^{\nu_n]}_{\nu_n'}
\left (\prod_{l=1}^n (\partial_{\mu_l} \Psi_{\nu_l})(\partial^{\mu_l} \Psi^{\nu_l'})\right)
\partial_t \Psi_k}
~
\begin{array}{c}
-\\
\\
\\
\\
\\
\\
\end{array}
~
\name{(\textrm{Venus})}{n\delta^{[\nu_1}_{\nu_1'} ... \delta^{\nu_n]}_{\nu_n'} 
 (\partial_t\Psi_{\nu_1})(\partial^{\mu_1} \Psi^{\nu_1'})
 \left (\prod_{l=2}^n (\partial_{\mu_l} \Psi_{\nu_l}) (\partial^{\mu_l} \Psi^{\nu_l'}) \right ) \partial_{\mu_1} \Psi_k.
 }\]
We split the term $(\textrm{Venus})$ into $\nu_1=k$ and $\nu_1\neq k$ as follows
\begin{multline}
(\textrm{Venus})=
\name{(\textrm{Tellus})}{
n\delta^{[k}_{\nu_1'} \delta^{\nu_2}_{\nu_2'}... \delta^{\nu_n]}_{\nu_n'} 
 (\partial_t\Psi_{k})(\partial^{\mu_1} \Psi^{\nu_1'})
 \left (\prod_{l=2}^n (\partial_{\mu_l} \Psi_{\nu_l}) (\partial^{\mu_l} \Psi^{\nu_l'}) \right ) \partial_{\mu_1} \Psi_k} \\
\name{(\textrm{Mars})}{
+n
\sum_{\nu_1\neq k}
\delta^{[\nu_1}_{\nu_1'} ... \delta^{\nu_n]}_{\nu_n'} 
 (\partial_t\Psi_{\nu_1})(\partial^{\mu_1} \Psi^{\nu_1'})
 \left (\prod_{l=2}^n (\partial_{\mu_l} \Psi_{\nu_l}) (\partial^{\mu_l} \Psi^{\nu_l'}) \right ) \partial_{\mu_1} \Psi_k.}
\end{multline}
$(\textrm{Mars})$ is identically zero, which the following argument shows: 
Because of the antisymmetrization over $\nu_1...\nu_n$, in every term in $(\textrm{Mars})$ $\nu_1,...,\nu_n$ will take every value $1,...,n$. 
Since $\nu_1\neq k$, it means that there is some $m>1$ such that $\nu_m = k$. 
Isolating the corresponding factor from the product, we get 
\[
(\textrm{Mars}) = 
n
\sum_{\nu_1\neq k}
\delta^{[\nu_1}_{\nu_1'} ... \delta^{\nu_n]}_{\nu_n'} 
 (\partial_t\Psi_{\nu_1})
 (\partial^{\mu_1} \Psi^{\nu_1'}) (\partial_{\mu_1} \Psi_k)
 (\partial_{\mu_m} \Psi_{k}) (\partial^{\mu_m} \Psi^{\nu_m'})
 \left (\prod_{l=2, l\neq m}^n (\partial_{\mu_l} \Psi_{\nu_l}) (\partial^{\mu_l} \Psi^{\nu_l'}) \right ) = 0
\]
because the factor $ (\partial^{\mu_1} \Psi^{\nu_1'}) (\partial_{\mu_1} \Psi_k)
 (\partial_{\mu_m} \Psi_{k}) (\partial^{\mu_m} \Psi^{\nu_m'})$, this is symmetric under the interchange $\nu_1'\leftrightarrow \nu_m'$, but the Kronicker-delta product is antisymmetric. 
Now consider  $(\textrm{Mercury})$.
As before, in every term, $\nu_1,...,\nu_n$ will take every value $1,...,n$. 
Thus, in each term of $(\textrm{Mercury})$, there will be an $m$ such that $\nu_m=k$, so we write
\begin{multline}
(\textrm{Mercury})
=
\sum_{m=1}^n \delta^{[\nu_1}_{\nu_1'} ... \delta^{k}_{\nu_m'} ... \delta^{\nu_n]}_{\nu_n'}
(\partial_{\mu_m} \Psi_{k})(\partial^{\mu_m} \Psi^{\nu_m'})
\left (\prod_{l=1 \neq m}^n (\partial_{\mu_l} \Psi_{\nu_l})(\partial^{\mu_l} \Psi^{\nu_l'})\right) (\partial_t \Psi_{k}) \\
=
\sum_{m=1}^n \delta^{[k}_{\nu_m'} \delta^{\nu_1}_{\nu_1'} ... \delta^{\nu_{m-1}}_{\nu_{m-1}'} \delta^{\nu_{m+1}}_{\nu_{m+1}'}... \delta^{\nu_n]}_{\nu_n'}
(\partial_{\mu_m} \Psi_{k})(\partial^{\mu_m} \Psi^{\nu_m'})
\left (\prod_{l=1 \neq m}^n (\partial_{\mu_l} \Psi_{\nu_l})(\partial^{\mu_l} \Psi^{\nu_l'})\right) (\partial_t \Psi_{k}).
\end{multline}
Now, renaming $\nu_m'\rightarrow \nu_1'$, $\nu_1\rightarrow \nu_2$, $\nu_1'\rightarrow \nu_2'$, and so on up to $\nu_{m-1}'\rightarrow \nu_{m}'$, and $\mu_m\rightarrow \mu_1$, $\mu_1\rightarrow \mu_2$, and so on up to $\mu_{m-1}\rightarrow \mu_m$, we get 
\begin{multline}
    (\textrm{Mercury})
    =
    \sum_{m=1}^n \delta^{[k}_{\nu_1'} \delta^{\nu_2}_{\nu_2'} ...  \delta^{\nu_n]}_{\nu_n'}
(\partial_{\mu_1} \Psi_{k})(\partial^{\mu_1} \Psi^{\nu_1'})
\left (\prod_{l=2}^n (\partial_{\mu_l} \Psi_{\nu_l})(\partial^{\mu_l} \Psi^{\nu_l'})\right) (\partial_t \Psi_{k}) \\
=
n
\delta^{[k}_{\nu_1'} \delta^{\nu_2}_{\nu_2'} ...  \delta^{\nu_n]}_{\nu_n'}
(\partial_{\mu_1} \Psi_{k})(\partial^{\mu_1} \Psi^{\nu_1'})
\left (\prod_{l=2}^n (\partial_{\mu_l} \Psi_{\nu_l})(\partial^{\mu_l} \Psi^{\nu_l'})\right) (\partial_t \Psi_{k}) = (\textrm{Tellus}).
\end{multline}
This, in turn, means $(\textrm{Mercury}) = (\textrm{Venus})$, which shows that $v^{\mu_1}_{\textrm{candidate}}$ is a solution to Equation (\ref{eq:appendix_advection_of_psi}). 
We have verified this calculation explicitly up to $n=d=5$, using symbolic mathematical software. 
In addition, it is straight-forward to show that $v^{\mu_1}_{\textrm{candidate}}$ is orthogonal to $\rho_{i_1...i_{d-n}}$ in the sense that $v^{i_k}_{\textrm{candidate}} \rho_{i_1...i_k...i_{d-n}}=0$ for all $k$. 
Identifying these as the $(d-n)$ necessary conditions to determine $\vec v^{(\Psi)}$, we get $\vec v_{(\Psi)} = \vec v_{\textrm{candidate}}$, and fix the gauge on $\vec v$ by $\vec v=\vec v_{(\Psi)}$.
which gives, finally, the closed expression for the velocity
\[
\boxed{
v^\mu
=
-n 
\frac{
 \delta^{[\nu_1}_{\nu_1'} ... \delta^{\nu_n]}_{\nu_n'} 
 (\partial_t\Psi_{\nu_1})(\partial^{\mu_1} \Psi^{\nu_1'})
 \prod_{l=2}^n (\partial_{\mu_l} \Psi_{\nu_l}) (\partial^{\mu_l} \Psi^{\nu_l'})
}{
\delta^{[\nu_1}_{\nu_1'} ... \delta^{\nu_n]}_{\nu_n'} 
\prod_{l=1}^n (\partial_{\mu_l} \Psi_{\nu_l})(\partial^{\mu_l} \Psi^{\nu_l'})
}
}
\]
While this derivation holds in general, we note that in the case of $n=d$, there is no contraction in getting to Equation \eqref{eq:velocity_after_contraction}, so the velocity can be equivalently written as 
\[
\textrm{Special case } n=d:  \quad
v^\mu = \frac{{J^{\mu}}}{\rho} = 
-n\frac{
\epsilon^{\mu_1...\mu_n}\tilde \epsilon^{\nu_1...\nu_n} (\partial_t \Psi_{\nu_1}) \prod_{l=2}^n (\partial_{\mu_l} \Psi_{\nu_l})
}{
\epsilon^{\mu_1...\mu_n}\tilde \epsilon^{\nu_1...\nu_n} \prod_{l=1}^n \partial_{\mu_l} \Psi_{\nu_l}
},
\]
which together give the expressions of the velocities in Equations (14-15) of the main article.
For the physically most interesting cases of $n\leq d\leq 3$, see Supplementary Figure \ref{fig:table_overview}.
\vspace{\columnsep}

\begin{figure*}
    \centering
    \includegraphics[]{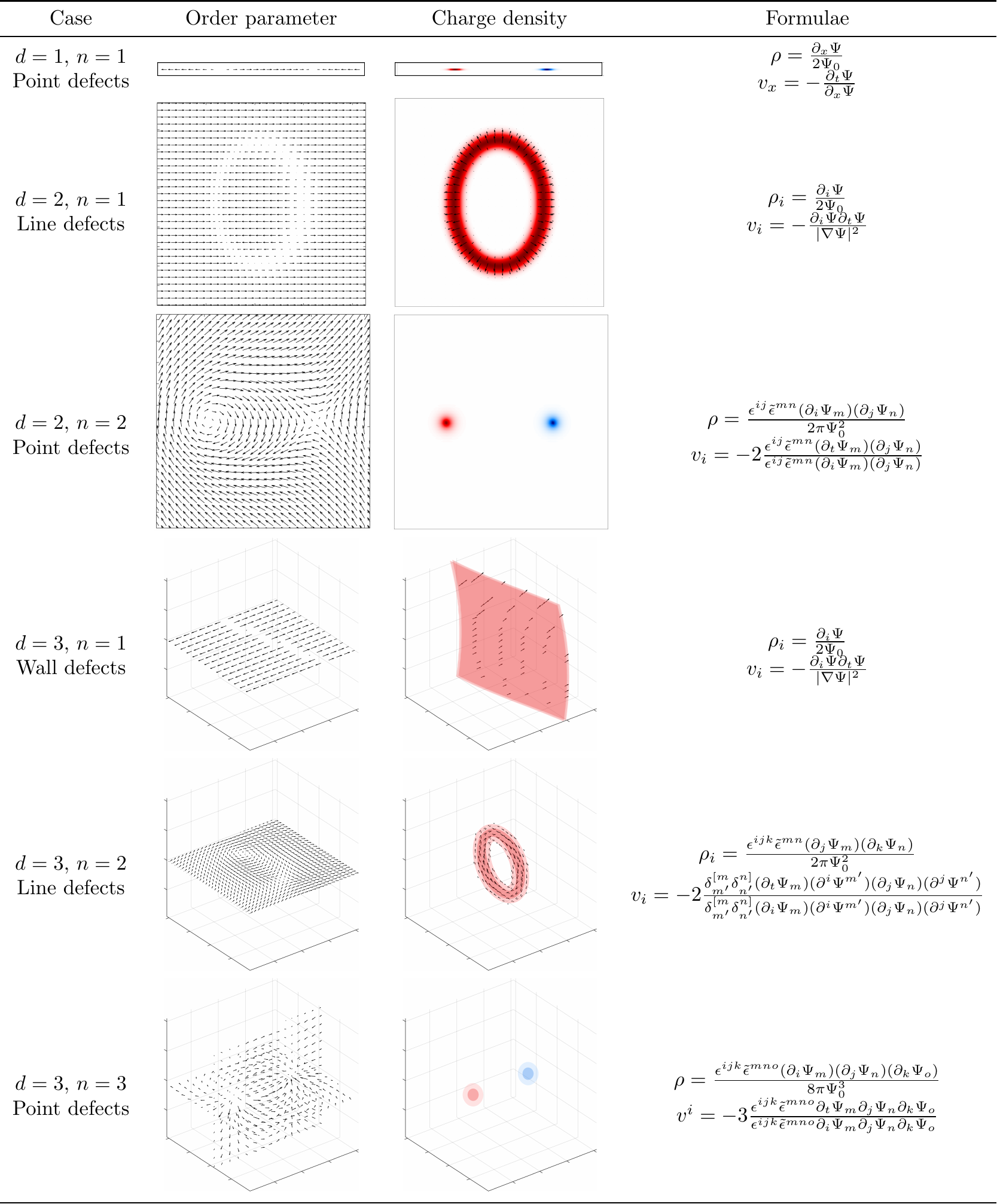}
    \caption{Examples of reduced defect field corresponding to stable topological defects in $O(n)$ models for $n=1,2,3$ in $d=1,2,3$. While the manuscript has centered on systems where $n=2$, the methodology can be applied to other cases, such as interfacial systems with $n=1$ with wall defects \cite{nguyenPhasefieldSimulationsViscous2010} or systems with $n=3$ such as the 3D Heisenberg model which features emergent magnetic monopoles as topological defects \cite{kanazawaDirectObservationStatics2020}.
    }
    \label{fig:table_overview}
\end{figure*}

\end{document}